\documentclass{article}
\usepackage{arxiv}
\renewcommand{\shorttitle}{}
\usepackage[utf8]{inputenc}
\usepackage{lscape}
\usepackage{subcaption}
\usepackage{amsmath}
\usepackage{cleveref}

\usepackage{amssymb}
\usepackage{amsthm}
\usepackage{bbm}
\usepackage{amsfonts}
\usepackage{dsfont}
\usepackage{color}
\usepackage{tcolorbox}
\usepackage{arydshln}
\usepackage[ruled,vlined]{algorithm2e}
\SetKwRepeat{Do}{do}{while}%
\usepackage{pifont}% http://ctan.org/pkg/pifont
\newcommand{\RN}[1]{%
	\textup{\uppercase\expandafter{\romannumeral#1}}%
}
\usepackage{graphicx}
\graphicspath{./home/marzsaee/Bureau/saeed/Thesis/Latex/}
\usepackage{subcaption}
\usepackage{listings}
\usepackage{multirow}
\usepackage{color} %red, green, blue, yellow, cyan, magenta, black, white
\definecolor{mygreen}{RGB}{28,172,0} % color values Red, Green, Blue
\definecolor{mylilas}{RGB}{170,55,241}

\usepackage[round]{natbib}

\newtheorem{theorem}{Theorem}[section]

\newtheorem{definition}{Definition}[section]
\newtheorem{assumption}{Assumption}[section]
\newtheorem{remark}{Remark}[section]
\newtheorem{lemma}{Lemma}[section]

\newcommand{\mynewwidth}{.45}
\newcommand{\mynewscale}{.45}

\newcommand{\price}{p}
\newcommand{\quoteIt}[1]{``#1''}

\newcommand{\1}{\textbf{1}}
\newcommand{\removed}[1]{}
\newcommand{\nonSelectedStructur}[1]{}

\newcommand{\Expect}{{\mathbb{E}}}
\renewcommand{\P}{{\mathbb{P}}}
\newcommand{\sProcess}{\psi}

\newcommand{\DRM}{{DRM}}
\newcommand{\SRM}{{SRM}}
\newcommand{\APPL}{{APPL}}
\newcommand{\AMZN}{{AMZN}}
\newcommand{\FB}{{FB}}
\newcommand{\JPM}{{JPM}}
\newcommand{\GOOGL}{{GOOGL}}
\newcommand{\DP}{{DP-DRM}}

\newif \iffinal
\finaltrue
\iffinal
\newcommand{\EDmodified}[1]{{#1}}
\newcommand{\EDcomments}[1]{{}}
\newcommand{\SMmodified}[1]{{#1}}
\newcommand{\SMcomments}[1]{{}}
\else
\newcommand{\EDmodified}[1]{{\color{red} #1}}
\newcommand{\EDcomments}[1]{{\EDmodified{Erick commented: #1}}}
\newcommand{\SMmodified}[1]{{\color{blue} #1}}
\newcommand{\SMcomments}[1]{{\SMmodified{Saeed commented: #1}}}
\fi

\title{Deep Reinforcement Learning for Equal Risk Pricing and Hedging under Dynamic Expectile Risk Measures}

\author{
  Saeed Marzban \\
  HEC Montr\'eal, Montr\'eal,\\
  H3T 2A7, Canada\\
  \texttt{saeed.marzban@hec.ca} \\
   \And
 Erick Delage \\
  HEC Montr\'eal, Montr\'eal,\\
  H3T 2A7, Canada\\
  \texttt{erick.delage@hec.ca} \\
  \And
  Jonathan Yumeng Li \\
  Telfer School of Management, University of Ottawa,\\
  Ottawa, Ontario K1N 6N5, Canada\\
  \texttt{jonathan.li@telfer.uottawa.ca} \\
}

\begin{document}

\maketitle

\begin{abstract}

Recently equal risk pricing, a framework for fair derivative pricing, was extended to consider dynamic risk measures. However, all current implementations either employ a static risk measure that violates time consistency, or are based on traditional dynamic programming solution schemes that are impracticable in problems with a large number of underlying assets (due to the curse of dimensionality) or  with incomplete asset dynamics information. In this paper, we extend for the first time a famous off-policy deterministic actor-critic deep reinforcement learning (ACRL) algorithm to the problem of solving a risk averse Markov decision process that models risk using a time consistent recursive expectile risk measure. This new ACRL algorithm allows us to identify high quality time consistent hedging policies (and equal risk prices) for options, such as basket options, that cannot be handled using traditional methods, or in context where only historical trajectories of the underlying assets are available. Our numerical experiments, which involve both a simple vanilla option and a more exotic basket option, confirm that the new ACRL algorithm can produce 1) in simple environments, nearly optimal hedging policies, and highly accurate prices, simultaneously for a range of maturities  2) in complex environments, good quality policies and prices using reasonable amount of computing resources; and 3) overall, hedging strategies that actually outperform the strategies produced using static risk measures when the risk is evaluated at later points of time.

\end{abstract}


\section{Introduction}

Derivative pricing remains to be a challenging problem in finance when the markets are incomplete and the derivatives are dependent on multiple underlying assets. The incompleteness of a market implies that the price of a derivative cannot be uniquely determined by the standard replication argument, as in such a market no self-financing hedging strategy exists that can perfectly replicate the payoffs of the derivative. Many mechanisms have been proposed for pricing in an incomplete market but most were developed from the perspective of a single trader. Unfortunately, a price that is set only according to one party's interest, e.g. a super-replication price that a seller may wish to charge, may not be acceptable to the buyer and thus does not represent a plausible transaction price. Recently, a new pricing scheme, known as Equal Risk Pricing (ERP), was proposed by \cite{guo2017equal} and further adapted to  convex risk measures in 
the work of \cite{marzban2020equal}.

The scheme of ERP is built upon the idea of modelling separately the risk exposure of the buyer and the seller of a derivative, and seeking a price that ensures that the risk exposure of both parties is the same 
under their respective optimal self-financing hedging strategy. The price generated from ERP thus has the merit of fairness to both parties. 
While ERP has its conceptual appeal, there remains a gap between its general construct and the actual implementation.  In particular, as shown in \cite{marzban2020equal}, great care must be taken to define properly how risk should be measured in a dynamic hedging setting in order to obtain hedging problems that are operationally meaningful and computationally solvable. The work of \cite{marzban2020equal} provides necessary analysis for solving the equal risk pricing and hedging problem based on dynamic programming (DP).  It is known however that DP suffers from the issue of the curse of dimensionality, which restricts the applicability of the results in \cite{marzban2020equal}. In addition, DP assumes the knowledge of a stochastic model that precisely captures the dynamics of the markets, which may not be available in  practice.

In the past decade, 
%In this paper, our goal is to provide an alternative approach for solving ERP problems, namely based on deep reinforcement learning (DRL). 
Deep Reinforcement Learning (DRL) has proven to be a powerful tool for solving dynamic optimization problems when the number of state variables is large and/or when no stochastic model is known for the underlying system dynamics.   In particular, the recent works of \cite{carbonneau2020equal}  and \cite{carbonneau2021deep} are the first that apply DRL to solve ERP problems and they demonstrate the possibility of pricing a broad range of over-the-counter options such as basket options. Unfortunately, the DRL approaches proposed in 
\cite{carbonneau2020equal}  and \cite{carbonneau2021deep} can only be used in settings where the risk is measured according to a static risk measure. This raises the serious issue that the hedging problem exploited by the ERP could be time inconsistent, i.e. the hedging decisions planned for future state of the world may not be considered optimal anymore once the state is visited.
%under this setting may not be optimal with respect to the scenarios that can happen afterwards. %Another way to put it is that the hedging decisions under this setting may depend on scenarios that one already known cannot happen in the future. 
The violation of time consistency implies that equal risk prices calculated based on static risk measures will assume a hedging policy that cannot be implemented in practice, and thus are optimistically biased. 
%is implausibility would make hedging decisions generated based on static risk measures difficult to implement in practice and thus also renders the derivative price resulting from the hedging decisions less justifiable. 
From a numerical perspective, employing a static risk mesure in ERP also limits the type of DRL algorithms that can be used to solve each party's hedging problem. Specifically, the authors of \cite{carbonneau2020equal}  and \cite{carbonneau2021deep} employ a policy optimization scheme, a.k.a. Actor-Only RL (AORL) algorithm (see \cite{williams1992simple} as an example of this method), while other approaches such as critic-only or actor-critic algorithms (such as \cite{mnih2015human} and \cite{silver2014deterministic} respectively) that rely on an equivalent DP formulation remain out of reach.
%The setting also has an implication on the use of DRL to solve the hedging problem. As the use of static risk measures would break the dynamic optimality principle in solving the hedging problem, \cite{carbonneau2020equal}  and \cite{carbonneau2021deep} resort to an Actor-Only RL (AORL) architecture for determining a policy network. This architecture relies on the simulation of a large number of trajectory paths in order to estimate risk and update a hedging policy network. It does not involve the design of a Q-update, i.e. the update of a state-action value function. The focus of this paper is different. 

In this paper, we seek to develop a DRL approach for solving a class of time-consistent ERP problems. It is known that to ensure time consistency, a dynamic risk measure should be employed to measure risk in a recursive fashion. In particular, motivated by the theory of coherent risk measures, which identifies expectile risk measures as the only elicitable coherent risk measures, we propose in this paper the use of dynamic expectile risk measures to formulate time consistent ERP problems. The dynamic nature of risk measures suggests the consideration of an Actor-Critic RL (ACRL) algorithm for solving the hedging problem. It turns out that the elicitability property of expectile risk measures facilitates greatly the design of a model-free ACRL algorithm. The convergence of this algorithm is also greatly improved due to the translation invariance property of the risk measures.
%, i.e. allowing a policy to be updated per sample, and some other property, namely translation invariance, allows the hedging problem to be recast into a form that can be solved more efficiently by ACRL. 

Overall, we may summarize the contribution of this paper as follows:
\begin{itemize}
    \item We present the first model-free DRL based algorithm for computing equal risk prices that rely on option hedging strategies that are time-consistent. To reinforce the importance of this contribution, we in fact demonstrate using a simple single asset two-period horizon option pricing problem how equal risk prices might suffer from an optimistic bias when static risk measures are used (as in \cite{carbonneau2020equal} and \cite{carbonneau2021deep}). A side benefit from pricing an option with maturity $T$ using dynamic risk measures will be that we will easily obtain equal risk prices for any other maturity $T'< T$.
    \item The ACRL algorithm that we propose is the first model-free DRL algorithm to naturally extend the famous off-policy deterministic actor-critic method presented in \cite{silver2014deterministic} to the risk averse setting. Unlike the ACRL proposed in \cite{Tamar2015:PGCRM} and \cite{huang2021convergence} for risk-averse DRL, which can employ up to five neural networks, our algorithm will only require two deep neural networks: a policy network (actor) and a Q network (critic). While our policy network will be trained following a stochastic gradient procedure similar to \cite{silver2014deterministic}, to the best of our knowledge we are the first to leverage the elicitability property (i.e. existence of a scoring function) of expectile risk measures and to propose a procedure for training the \quoteIt{risk-to-go} Q network that is also based on stochastic gradient descent.
\item We perform a comprehensive evaluation of the training efficiency, quality of option hedging strategies, and quality of equal risk prices obtained using our ACRL algorithm on a synthetic multi-asset geometric Brownian motion market model. In the simple case of vanilla option pricing, we provide empirical evidence that ACRL provides nearly optimal hedging policies, and highly accurate prices, simultaneously for a range of maturities. The latter is in sharp contrast with approaches, such as in \cite{carbonneau2021deep}, that employ time inconsistent risk measures and produce investment strategies that are visibly outperformed by the ACRL strategy in terms of the risk measured as time to maturity reduces. This phenomenon is also observed, although less prominently, in the context of a basket option over 5 underlying assets, where good quality policies and prices are obtained using our ACRL algorithm using a reasonable amount of computing resources.

%    \item We perform a comprehensive evaluation of the training efficiency, quality of equal risk prices, and quality of option hedging strategies obtained using our ACRL, which employs a dynamic risk model (\DRM{}), and a similar AORL approach as used in \cite{carbonneau2021deep}, which employs a static risk model (\SRM{}), on a synthetic multi-asset geometric Brownian motion market model. We first confirm that both implementations of DRL successfully achieve high degree precision on a vanilla option pricing problem, for which we have access to exact pricing and hedging strategy using DP. We then demonstrate empirically that exploiting translation invariance to provide immediate feedback to the policy optimization scheme in the DRM approach significantly improves the speed of convergence and quality of the trained network. Finally,   \EDcomments{Talk about different maturities and Basket option experiments}
\end{itemize}

\EDcomments{OLD TEXT:
We should further elaborate these points by discussing first the ACRL architecture developed in this paper. The architecture is built upon the formulation of Q-value dynamic equations for the expectile-based hedging problems, and it consists of two networks, a policy network (actor) and a Q network (critic). The novelty lies in the design of an algorithm used to update the two networks based on stochastic gradient. In particular, our algorithm may be considered as an extension of the off-policy deterministic actor-critic method. Namely, following similar arguments made in Degris et al (2014), the algorithm updates the policy network based on off-policy deterministic policy gradient. Updating the Q network requires additional care, as the hedging problem is evaluated in terms of risk rather than expected value commonly applied in a RL problem. By leveraging on the elicitability property of expectile risk measures, which implies  that the risk measure can always be calculated as the optimal solution with respect to certain score function, we show that the algorithm can update the Q network using stochastic gradient calculated based on the score function. 
To the best of our knowledge, our algorithm is the first model-free actor-critic algorithm for solving problems employing dynamic risk measures. 
%We observe that this ACRL has a particularly good convergence behaviour when there is an immediate reward following each action. While the hedging problem in ERP is only concerned about the cumulative wealth and the payoff of a derivative that occur at the very end, the translation invariance property of expectile risk measures allows the cumulative wealth to be re-expressed as a sequence of immediate rewards that need to be optimized over time. Indeed, throughout our experiments, noticeably stronger convergence behaviour is found for the hedging problem formulated in terms of immediate rewards than the one formulated based on cumulative wealth. The effectiveness of our ACRL for finding a good policy is further demonstrated in the numerical section. 

We should mention that expectile risk measure has also been considered in the context of Distributional Reinforcement Learning...

({\color{red}Findings from the numerical section})
In the numerical section of this paper we first investigate a vanilla option where our purpose is twofold. First, we want to show the Q-function is precise enough to be used for the sake of pricing based on ERP framework. This is performed by comparing the results of this model with a grid-based DP model that can be trusted to provide a baseline for the value function. The results show that the Q-function can approximate the true value function when the policies are coming from the ACRL model, however, if the policies are chosen according to an \SRM{} model, the quality of this approximation diminishes. Second, we demonstrate the benefits of having a model that provides a time consistent solution in practice. The numerical results support our claim that a time consistent model can be used for training a model over options with long maturities and then use the trained model to hedge and price options with shorter maturities. We also show the benefits of formulating the option pricing problem  such that the immediate rewards are included in the RL setting. More precisely, we show this transformation will improve the convergence ability of the model significantly. Finally, we focus on basket options where we demonstrate the main purpose of extending the ERP model to using RL for pricing and hedging, which is improving the scalability. The results in this section follow our previous results in the case of vanilla option where the time consistent solution is able of outperforming the static time inconsistent solution.}

The rest of this paper is organized as follows. Section \ref{sec:ERP} introduces equal risk pricing and illustrates using a simple two-period pricing problem the  practical issues related to using static risk measures for option hedging and pricing. Section \ref{sec:EPR_DP_Alg} adapts the ERP framework to the case of a dynamic expectile risk measure and proposes the new ACRL algorithm. Finally, Section \ref{sec:Experiments} presents and discusses our numerical experiments.

\section{Equal risk pricing and hedging under coherent risk measures}\label{sec:ERP}

In this section, we provide a brief overview of ERP under coherent risk measures based on the recent work of \cite{marzban2020equal}. We pay particular attention to the issue of time (in)consistency by presenting an example that demonstrates numerically that employing a time-inconsistent static risk measure can lead to an under-evaluation of the risk to which each party are actually exposed in practice.
%the sub-optimality of a time inconsistent hedging strategy. 

\subsection{ERP under coherent risk measures}
The problem of ERP can be formalized as follows. Consider a frictionless market, i.e. no transaction cost, tax, etc, that contains $m$ risky assets, and a risk-free bank account with zero interest rate. Let $\mathbf{S}_t : \Omega  \rightarrow \mathbb{R}^m$ denote the values of the risky assets adapted to a filtered probability space $ (\Omega,\mathcal{F},\mathbb{F}:=\{\mathcal{F}_t\}_{t=0}^T,\mathbb{P})$, i.e. each $\mathbf{S}_t$ is $\mathcal{F}_t$ measurable. It is assumed that $\mathbf{S}_t$ is a locally bounded real-valued semi-martingale process and that the set of equivalent local martingale measures is non-empty (i.e. no arbitrage opportunity). The set of all admissible self-financing hedging strategies with the initial capital $ \price_0 \in \mathbb{R}$ is shown by $ \mathcal{X}(\price_0)$:
\begin{equation}\nonumber
\mathcal{X}(\price_0)=\left\{X:\Omega\rightarrow \mathbb{R}^T \left| \exists \{\boldsymbol{\xi}_t\}_{t=0}^{T-1}, \quad X_t=\price_0+\sum_{t'=0}^{t-1} \boldsymbol{\xi}_{t'}^\top \Delta \mathbf{S}_{t'+1} , \quad \forall t=1,\dots,T \right. \right\}\,,
\end{equation}
where $ \Delta \mathbf{S}_{t+1} :=\mathbf{S}_{t+1}-\mathbf{S}_{t} $, the hedging strategy $\boldsymbol{\xi}_t \in \mathbb{R}^m$ is a vector of random variables adapted to the filtration $\mathbb{F}$ and captures the number of shares of each of the risky assets held in the portfolio during the period $[t,\;t+1]$, $\boldsymbol{\xi}_{t}^\top 
\Delta \mathbf{S}_{t'+1}$ is the inner product of the two random vectors, and  $ X_{t} $ is the accumulated wealth. 

Let $F(\{\mathbf{S}_t\}_{t=1}^T)$ denote the payoff of a derivative. Throughout this paper, we assume $F(\{\mathbf{S}_t\}_{t=1}^T)$ admits the formulation of $F(\mathbf{S}_T, \mathbf{Y}_T)$ where $\mathbf{Y}_t$ is an auxiliary fixed-dimensional stochastic process that is $\mathcal{F}_t$-measurable. This class of payoff functions is common in the literature, (see for example \cite{bertsimas2001hedging} and \cite{marzban2020equal}). The problem of ERP is defined based on the following two hedging problems that seek to minimize the risk of hedging strategies, one is for the writer and the other is for the buyer of the derivative: 
%\begin{subequations}\label{equation:riskMeasures}
	\begin{align}
	\mbox{(Writer)}\quad\quad&\varrho^w(\price_0)=\inf_{X\in\mathcal{X}(\price_0)} \rho^{w}(F(\mathbf{S}_{T},\mathbf{Y}_{T})-X_{T}) \label{equation:riskMeasures1}\\
	\mbox{(Buyer)}\quad\quad&\varrho^b(\price_0)=\inf_{X\in\mathcal{X}(-\price_0)} \rho^{b}(-F(\mathbf{S}_{T},\mathbf{Y}_{T})-X_{T})\,,\label{equation:riskMeasures2}
	\end{align}
%\end{subequations}
where $\rho^w$ and $\rho^b$ are two risk measures that capture respectively the writer and the buyer's risk aversion. In words, equation \eqref{equation:riskMeasures1} describes a writer that is receiving $ \price_0 $ as the initial payment and implements an optimal hedging strategy for the liability captured by $F(\mathbf{S}_{T},\mathbf{Y}_{T})$. On the other hand, in \eqref{equation:riskMeasures2} the buyer is assumed to borrow $\price_0 $ in order to pay for the option and then to manage a portfolio that will minimize the risks associated to his final wealth $F(\mathbf{S}_{T},\mathbf{Y}_{T})+X_{T}$. With equations \eqref{equation:riskMeasures1} and \eqref{equation:riskMeasures2}, ERP defines a fair price $p_0^*$ as the value of an initial capital that leads to the same risk exposure to both parties, i.e. 
$$\rho^w(p_0^*) = \rho^b(p_0^*).$$

Motivated by the theory of coherent risk measures (\cite{artzner1999coherent}), \cite{marzban2020equal} study the ERP problem by imposing the property of coherency to the risk measures $\rho^w$ and $\rho^b$. Namely, a risk measure is said to be coherent if it satisfies the following five conditions:
\begin{itemize}
	\item Monotonicity: if $ X \le Z \quad a.s.$ then $ \rho(X) \le \rho(Z) $
	\item Subadditivity: $ \rho(X+Z) \le \rho(X) + \rho(Z) $
	\item Positive homogeneity: If $ \lambda \ge 0 $, then $ \rho(\lambda X)=\lambda \rho(X) $
	\item Translation invariance: If $ m \in \mathbb{R} $, then $ \rho(X+m)=\rho(X)+m $
	\item Normalized risk: $\rho(0)=0$.
\end{itemize}

It is well known that Value-at-Risk (VaR), a risk measure commonly applied in financial risk management, is not coherent, whereas its convex counterpart, namely Conditional Value-at-Risk (CVaR) is coherent. The application of CVaR in ERP can be found for example in \cite{carbonneau2020equal}. As one of the key results in ERP, \cite{marzban2020equal} establishes that an equal risk price $p_0^*$ can actually be found by solving the writer and buyer's hedging problem with no initial payment, i.e. \eqref{equation:riskMeasures1} and \eqref{equation:riskMeasures2}, separately. Namely, it can be calculated by the following result.

\begin{theorem} Let $\rho^w$ and $\rho^b$ be two coherent risk measures. In the case where the equal risk price $p_0^*$ exists, it can be calculated by 
\[\price_0^*=(\varrho^w(0)-\varrho^b(0))/2\,,\]
when $\infty>\varrho^w(0)\geq \varrho^b(0)>-\infty$.
\end{theorem}

%\begin{definition}\label{def:erp}
%[See \cite{marzban2020equal}] (Equal risk price) Given that both the writer's risk measure, $ \rho^w$, and buyer's risk measure $ \rho^b $ are coherent risk measures, the equal risk price is defined as the unique $\price_0^* $ that satisfies
%\begin{equation}\label{eq:ERP}
% \varrho^w(\price_0^*)=\varrho^b(\price_0^*) \in \mathbb{R}\,,
%\end{equation}
%when such a unique price exists.
%\end{definition}

\subsection{The issue of time inconsistency}\label{sec:simple_example}
As briefly mentioned in the introduction, measuring risk in a dynamic setting requires additional care. The use of a coherent risk measure, without any further adaptation to a dynamic setting, can lead to solutions that suffer from the issue of time inconsistency. The goal of this section is to carefully demonstrate this point by presenting a numerical example that quantifies the impact of time inconsistency. 
%This offers the motivation for the rest of this work. 
Our demonstration is inspired by the work of \cite{rudloff2014time}, where the impact of time inconsistency is discussed in a portfolio management problem. Here, we present an example based on a vanilla option hedging problem. 

%This section elaborates on the issue of time inconsistency of multistage policies following the work of \cite{rudloff2014time} in which time consistency is defined as: ``\textit{an optimal policy is time consistent if and only if the future planned decisions are actually going to be implemented}". 

In this example, we consider a stock price process modelled by a simple two-stage trinomial tree. Specifically, the horizon spans $t \in \{0,1,2\}$ and the probability space $(\Omega,\mathcal{F},\mathbb{P})$ is such that $\Omega = \{\omega_i\}_{i=1}^9$,  %\mathcal{F}_1:=\sigma(\{(\omega_1,\omega_2,\omega_3),(\omega_4,\omega_5,\omega_6),(\omega_7,\omega_8,\omega_9)\})$
$\mathcal{F}_1:=\sigma(\{\{\omega_i\}_{i=1}^3,\{\omega_i\}_{i=4}^6,\{\omega_i\}_{i=7}^9\})$, and all outcomes are equiprobable. The market contains a risk-free asset (with a risk-free rate of zero) and a risky asset $S$ which are used to hedge a vanilla at-the-money call option on $S_2$ with strike price $K:=S_0$. The details of the price process is shown in Table \ref{tbl:simpleExampleStaticDynamic}. For simplicity, we set the initial capital for hedging to zero %dollar and short-selling is allowed. $X_{t}(\omega)$ is the accumulated wealth at time $t$, and $\zeta_{j,t}(\omega)$ denotes the dollar value invested in asset $j$ at time $t$, and both are measurable with respect to the filtration $\mathcal{F}$.
and employ a $\mbox{CVaR}_{60\%}$ risk measure for hedging.

%. Using the minimization representation of CVaR, we can formulate the hedging problem as a linear stochastic program. Recall the following definition of ${\rm CVaR}_{\alpha}$:
%\begin{equation}
%    \text{CVaR}_\alpha(Y) = \min_{c \in \mathbb{R}} \mathbb{E} \left[c+\frac{1}{1-\alpha}(Y-c)\mathbb{I}_{\{Y>c\}}\right].
%\end{equation}

 When hedging the call-option using a static CVaR measure, the writer of the option solves the %Then, a hedging strategy can be obtained by solving the 
 following two-period optimization model:
%  \begin{subequations}\label{eq:simple_example1}
%  \begin{eqnarray}\label{eq:simple_example1}
% \min_{\zeta_0,\zeta_1,X_1,X_2} \quad&&\mbox{CVaR}_\alpha((S_{2}^1(\omega)-K)^+-X_2(\omega))\\
% \mbox{subject to}&& X_{1}(\omega) = (S_1(\omega)-S_0)\zeta_0, \quad\forall \omega \in \Omega\\
% &&X_2(\omega) = X_1(\omega)+(S_2(\omega)-S_1(\omega))\zeta_1(\omega), \quad\forall \omega \in \Omega
% \end{eqnarray}
% \end{subequations}
 %\begin{subequations}
 \begin{eqnarray}\label{eq:simple_example1}
\min_{\xi_0,\xi_1} \quad&&\mbox{CVaR}_{60\%}((S_{2}(\omega)-K)^+-(S_1(\omega)-S_0)\xi_0-(S_2(\omega)-S_1(\omega))\xi_1(\omega))
\end{eqnarray}
%\end{subequations}
 where $(y)^+:=\max(y,0)$ and $K:=S_0$. The optimal solution of this problem will prescribe purchasing 0.93 shares of the risky asset at time 0, i.e. $\xi_0 = 0.9341$, using money borrowed at the risk-free rate (see Table \ref{tbl:simpleExampleStaticDynamic} for the optimal shares to hold at $t=1$). The resulting $\mbox{CVaR}_{60\%}$ is $26.36$, implying that if the writer charges the buyer with a price above $26.36$, the writer would consider the price being sufficient to  cover the hedged risk of this call option.

Note that in the risk averse hedging problem \eqref{eq:simple_example1}, it is not clear what motivates the writer of the option to implement the prescribed hedging strategy once new information is revealed at time $t=1$. In particular, he/she might be curious to compare the prescribed strategy with the strategy that minimizes the CVaR from the new perspective at $t=1$, i.e., the following hedging problem:
 \begin{eqnarray}\label{eq:simple_example2}
\min_{\bar{\xi}_1} \quad&&\mbox{CVaR}_{\bar{\alpha}_1}((S_{2}(\omega)-K)^+-(S_1(\omega)-S_0)\xi_0^*-(S_2(\omega)-S_1(\omega))\bar{\xi}_1(\omega)|\mathcal{F}_1)\,,
\end{eqnarray}
where $\bar{\alpha}_1:=60\%$ and where $\xi_0^*=0.9341$, i.e. the optimal first stage solution in \eqref{eq:simple_example1}.

Table \ref{tbl:simpleExampleStaticDynamic} presents the optimal conditional hedging strategy $\bar{\xi}_1^*$ as a function of the information revealed by $\mathcal{F}_1$. While it does appear that $\bar{\xi}_1^*$ agrees with $\xi_1^*$ when $\omega\in\{\omega_i\}_{i=1}^3$, the investment in the risky asset ends up significantly reduced in the other two sets of outcomes. More importantly, we established that in order to motivate the prescribed hedging strategy $\xi_1^*$, the risk aversion level used in problem \eqref{eq:simple_example2} would need to be in the range of $[0.4580,\;0.4585]$, when $\omega\in\{\omega_i\}_{i=4}^6$, or $[0.1992,\;0.2]$, when $\omega\in\{\omega_i\}_{i=7}^9$. This confirms that $\xi^*$ is likely to be perceived as overly risky given the information revealed at time $t=1$.
Ultimately, in the likely case where the writer decides to replace $\xi_1^*$ with $\bar{\xi}_1^*$, one can establish that the overall exposition to risk from the perspective of $t=0$ should have rather been estimated to 27.94 instead of 26.36. This implies that employing a static risk measure here underestimated the necessary coverage capital by $6\%$.

While this issue of time consistency has been discussed significantly in the recent years, a common approach to overcome it is to employ a so-called dynamic risk measure as will be done in the following section. In the context of this example, this would reduce to replacing problem \eqref{eq:simple_example1} with:
\begin{eqnarray}\label{eq:simple_example3}
\min_{\xi_0,\xi_1} \quad&&\mbox{CVaR}_{\alpha}(\;\mbox{CVaR}_{\alpha}((S_{2}(\omega)-K)^+-(S_2(\omega)-S_1(\omega))\xi_1(\omega)-(S_1(\omega)-S_0)\xi_0|\mathcal{F}_1(\omega))\;)\,,
\end{eqnarray}
where $\alpha$ can be chosen to characterize the right level of risk aversion for the \quoteIt{dynamic conditional value-at-risk measure}.  This formulation ensures that the prescribed policy at time $t=1$ remains optimal (according to problem \eqref{eq:simple_example2}) at the moment where it is actually implemented thus preventing the necessary coverage capital from being under estimated.

\begin{table}[h]
\caption{Example of a time inconsitent hedging strategy obtained from employing a static risk measure. $\xi^*$ is obtained by solving problem \eqref{eq:simple_example1}, $\bar{\alpha}_1$ is the risk aversion level that motivates $\xi_1^*$ at $t=1$, $\bar{\xi}_1^*$ is the actual investment prescribed by $\mbox{CVaR}_{60\%}$ at $t=1$.}
\begin{center}
    \begin{tabular}{c|ccc|ccc|c}
   Atoms & \multicolumn{3}{c|}{Price process} & \multicolumn{3}{c|}{Time inconsistent}  & \multicolumn{1}{c}{Optimal conditional}\\
    of $\mathcal{F}_1$ & \multicolumn{3}{c|}{} & \multicolumn{3}{c|}{hedging strategy}  & \multicolumn{1}{c}{hedging strategy}\\
    \cline{2-8}
    \vspace{-.3cm}
    &&&&&&&\\
     &	$S_{0}(\omega)$	&	$S_{1}(\omega)$	&	$S_{2}(\omega)$ 	&	
    $\xi^*_{0}$ 	&     $\xi^*_{1}(\omega)$ 	& $\bar{\alpha}_1(\xi^*)$ &	 $\bar{\xi}^*_{1}(\omega)$ \\
    \hline
    $\omega\in\{\omega_i\}_{i=1}^3$ 	&	100	&	150	&	\{270,150,75\}	& 0.9341	& 0.8718 & [0.4580,1.0000]	&	0.8718	\\
    $\omega\in\{\omega_i\}_{i=4}^6$ 	&	100	&	100	&	\{180,100,50\}	& 0.9341	& 0.7665	& [0.4580,0.4585] &	0.6154	\\
    $\omega\in\{\omega_i\}_{i=7}^9$ 	&	100	&	80	&	\{120,80,64\}	& 0.9341	& 0.5000	& [0.1992,0.2000] &	0.3571	\\
    
    \end{tabular}
    \label{tbl:simpleExampleStaticDynamic}
\end{center}
\end{table}

\section{ERP under dynamic expectile risk measure and an actor-critic algorithm} \label{sec:EPR_DP_Alg}

While time consistent ERP problems can be formulated by employing dynamic risk measures and be calculated, in principle, by solving a set of dynamic programming (DP) equations (\cite{marzban2020equal}), there remains the challenge of determining which dynamic risk measure one should employ and how these equations might be solved in high dimension, i.e. multiple underlying assets. In this section, we address the two issues by first motivating the use of dynamic expectile risk measures to formulate time consistent ERP hedging problems and then presenting a Deep Reinforcement Learning approach (DRL) for approximately solving this problems.

\subsection{Dynamic expectile risk measures and DP equations} \label{sec:EPR_DP}
Expectile has been proposed in the recent literature (see \cite{bellini2015:elicitable}) as a replacement of VaR and CVaR given that it is not only coherent but also elicitable. It is known that VaR is not coherent but is elicitable, whereas CVaR is coherent but is not elicitable. A risk measure is said to be elicitable if it can be expressed as the minimizer of a certain scoring function, and this property is found to be critical in practice due to the need of  backtesting \citep{chen2018exactitude}. In fact, expectile is the only elicitable coherent risk measure. Recall the following definition of expectile.

\begin{definition}\label{def:expectile}
%(See \cite{gneiting2011making})
(\cite{RMexpectile2017})
The $\tau-$expectile of a random liability $X$ is defined as:
\[\bar{\rho}(X):= \arg\min_{q}\,\tau\mathbb{E}\left[(q-X)_{+}^2\right]+(1-\tau)\mathbb{E}\left[(q-X)_{-}^2\right]\,.\]
\end{definition}

Like CVaR, expectile covers at one extreme the case of risk-neutrality, i.e. with $\tau = 1/2$, and at the other extreme the case of converging towards the worst-case risk, i.e. as $\tau \rightarrow 1$. Thus, expectile also allows for modelling a wide spectrum of risk aversion. Using expectile as the basis, we define its dynamic version as follows. 
\begin{definition}
A dynamic recursive expectile risk measure takes the form:
\[\rho(X):=\bar{\rho}_0(\bar{\rho}_1(\dots\bar{\rho}_{T-1}(X)))\,,\]
where each $\bar{\rho}(\cdot)$ is an expectile risk measure that employs the conditional distribution based on $\mathcal{F}_t$. Namely,
\[\bar{\rho}_t(X_{t+1}):=\arg\min_{q}\,\tau\mathbb{E}\left[(q-X_{t+1})_{+}^2|\mathcal{F}_t\right]+(1-\tau)\mathbb{E}\left[(q-X_{t+1})_{-}^2|\mathcal{F}_t\right]\]
where $X_{t+1}$ a random liability measureable on $\mathcal{F}_{t+1}$.
\end{definition}

We apply dynamic expecile risk measures to formulate the two hedging problems in ERP. By further imposing the following assumption that there exists a sufficient statistic process $\sProcess_t$ such that $\{(\mathbf{S}_t,\mathbf{Y}_t,\sProcess_t)\}_{t=0}^T$ satisfies the Markov property, we can obtain compact dynamic equations for them. 

\begin{assumption}\label{ass:Markov}[Markov property]
There exists a sufficient statistic process $\sProcess_t$ adapted to $\mathbb{F}$ such that $\{(\mathbf{S}_t,\mathbf{Y}_t,\sProcess_t)\}_{t=0}^T$ is a Markov process relative to the filtration $\mathbb{F}$. Namely, $\P((\mathbf{S}_{t+s},\mathbf{Y}_{t+s},\sProcess_{t+s})\in \mathcal{A}|\mathcal{F}_t) = \P((\mathbf{S}_{t+s},\mathbf{Y}_{t+s},\sProcess_{t+s})\in \mathcal{A}|\mathbf{S}_t,\mathbf{Y}_t,\sProcess_t)$ for all $t$, for all $s\geq 0$, and all sets $\mathcal{A}$.% in the Borel $\sigma$-algebra.
\end{assumption}

In particular, based on Proposition 3.1 and the examples presented in section 3.3 of \cite{marzban2020equal}, together with the fact that both $\rho^w$ and $\rho^b$ are dynamic recursive expectile risk measures, the Markovian assumption allows us to conclude that the ERP can be calculated using two sets of dynamic programming equations. Specifically, on the writer side, we can define
\[V^w_T(\mathbf{S}_T,\mathbf{Y}_T,\sProcess_T):= F(\mathbf{S}_T,\mathbf{Y}_T)\,,\]
and recursively
\[V^w_t(\mathbf{S}_t,\mathbf{Y}_t,\sProcess_t):= \inf_{\boldsymbol{\xi}_t} \bar{\rho}(-\boldsymbol{\xi}_t^\top \Delta \mathbf{S}_{t+1} + V_{t+1}^w(S_t+\Delta \mathbf{S}_{t+1},\mathbf{Y}_t+\Delta Y_{t+1},\sProcess_{t+1})|\mathbf{S}_t,\mathbf{Y}_t,\sProcess_t)\,,\]
where $\bar{\rho}(\cdot|\mathbf{S}_t,\mathbf{Y}_t,\sProcess_t)$ is the expectile risk measure that uses $\mathbb{P}(\cdot|\mathbf{S}_t,\mathbf{Y}_t,\sProcess_t)$. This leads to considering $\varrho^w(0)=V^w_0(\mathbf{S}_0,\mathbf{Y}_0,\sProcess_0)$. On the other hand, for the buyer we similarly define:
\[V^b_T(\mathbf{S}_T,\mathbf{Y}_T,\sProcess_T):= -F(\mathbf{S}_T,\mathbf{Y}_T)\,,\]
and
\[V^b_t(\mathbf{S}_t,\mathbf{Y}_t,\sProcess_t):= \inf_{\boldsymbol{\xi}_t} \bar{\rho}(-\boldsymbol{\xi}_t^\top \Delta \mathbf{S}_{t+1} + V_{t+1}^b(\mathbf{S}_t+\Delta \mathbf{S}_{t+1},\mathbf{Y}_t+\Delta \mathbf{Y}_{t+1},\sProcess_{t+1})|\mathbf{S}_t,\mathbf{Y}_t,\sProcess_t)\,,\]
with $\varrho^b(0)=V^b_0(\mathbf{S}_0,\mathbf{Y}_0,\sProcess_0)$. The following lemma summarizes how DP can be used to compute ERP.

\begin{lemma}
Under Assumption \ref{ass:Markov}, the ERP that employs dynamic recursive expectile riks measure can be computed as: $p_0^*=(V^w_0(\mathbf{S}_0,\mathbf{Y}_0,\sProcess_0)-V^b_0(\mathbf{S}_0,\mathbf{Y}_0,\sProcess_0))/2$.
\end{lemma}

\subsection{A novel Expectile-based actor-critic algorithm for ERP}\label{sec:ACRL}

In this section, we formulate each option hedging problem as a Markov Decision Process (MDP) denoted by $(\mathcal{S},\mathcal{A},r,P)$. In this regard, the agent (i.e. the writer or buyer) interacts with a stochastic environment by taking an action $a_t \equiv \boldsymbol{\xi}_t \in [-1,1]^m$ after observing the state $s_t \in \mathcal{S}$, which includes $\mathbf{S}_t$, $\mathbf{Y}_t$, and $\sProcess_t$. Note that to simplify exposition, in this section we drop the reference to the specific identity (i.e. $w$ or $b$) of the agent in our notation. 
The action taken at each time $t$ results in the immediate stochastic reward that takes the shape of the immediate hedging portfolio return, i.e. $r_{t}(s_t,a_t,s_{t+1}) := \boldsymbol{\xi}_{t}^\top\Delta \mathbf{S}_{t+1}$ when $t<T$ and otherwise of the option liability/payout $r_{T}(s_T,a_T,s_{T+1}) := F(\mathbf{S}_T,\mathbf{Y}_T)(1-2\cdot\1\{\mbox{agent$=$writer}\})$, which is insensitive to $s_{T+1}$. Finally, the Markovian exogeneous dynamics described in Assumption \ref{ass:Markov} are modeled using $P$ as $P(s_{t+1}|s_t,a_t)=\mathbb{P}(\mathbf{S}_{t+1},\mathbf{Y}_{t+1},\sProcess_{t+1}|\mathbf{S}_{t},\mathbf{Y}_{t},\sProcess_{t})$. Overall, each of the two 
%In the context of actor-critic reinforcement learning, the 
\SMmodified{dynamic} derivative hedging problems presented in Section \ref{sec:EPR_DP} reduce to a version of the following risk averse reinforcement learning problem:
\[\varrho(0)=V_0(\mathbf{S}_0,\mathbf{Y}_0,\sProcess_0)=\min_{\pi}Q_0^\pi(\bar{s}_0,\pi_0(\bar{s}_0))\;,\]
where $\bar{s}_0:=(\mathbf{S}_0,\mathbf{Y}_0,\sProcess_0)$ is the initial state in which the option is priced while
\[Q_t^\pi(s_t,a_t):=\SMmodified{\bar{\rho}}(-r(s_t,a_t,s_{t+1})+Q_{t+1}^\pi(s_{t+1},\pi_{t+1}(s_{t+1}))|\SMmodified{s_t})\;,\]
and $Q_T^\pi(s_T,a_T):=r_T(s_T,a_T,s_{T})$. By the nature of the MDP, which exploits the interchangeability property, we have that%The Bellman equation associated with 
\[\varrho(0)=Q_0^{\pi^*}(\bar{s}_0,\pi_0^*(\bar{s}_0))\;,\]
where
\[\pi_t^*(s_t)\in\arg\min_{a_t} \SMmodified{\bar{\rho}}(-r(s_t,a_t,s_{t+1})+Q_{t+1}^{\pi^*}(s_{t+1},\pi_{t+1}^*(s_{t+1}))|\SMmodified{s_t})\;=\;\arg\min_{a_t} Q_t^{\pi^*}(s_t,a_t)\;.\]
\EDcomments{This is actually short for:
\[\arg\min_{\pi}Q_0^\pi(\bar{s}_0,\pi_t(\bar{s}_0))\supseteq\{\pi:\pi_t(s_t)\in\arg\min_{a_t} \rho(-r(s_t,a_t,s_{t+1})+Q_{t+1}^{\pi^*}(s_{t+1},\pi_{t+1}^*(s_{t+1})))\}\]
}
Hence, $\pi^*$ must also minimize:
\begin{equation}
\pi^*\in\arg\min_{\pi} \Expect_{\substack{\tilde{t}\sim \{0,\dots,T-1\}\\s_{t+1}\sim P(\cdot|s_t,\bar{\pi}_t(s_t))}}[ Q_{\tilde{t}}^{\pi}(s_{\tilde{t}},\pi_{\tilde{t}}(s_{\tilde{t}}))]\label{eq:dpgobjective}
\end{equation}
where $s_0:=\bar{s}_0$ and $\tilde{t}$ is uniformly drawn from the range $\{0,\dots,T-1\}$, and where $\bar{\pi}$ is an arbitrary reference policy\footnote{In fact, given that $s_t$ is entirely exogeneous, the distribution of $s_{t+1}$ is unaffected by $\bar{\pi}$ in our option hedging problem.}.

\EDcomments{Again, the expression above is short for:
\[\arg\min_{\pi}Q_0^\pi(\bar{s}_0,\pi_t(\bar{s}_0))\supseteq\arg\min_{\pi} \Expect_{\substack{\tilde{t}\sim \{0,\dots,T-1\}\\s_{t+1}\sim P(\cdot|s_t,\bar{\pi}_t(s_t))}}[ Q_{\tilde{t}}^{\pi}(s_{\tilde{t}},\pi_{\tilde{t}}(s_{\tilde{t}}))]\]
}

In the context of a deep reinforcement learning approach, we can employ a procedure based on off-policy deterministic policy gradient \citep{silver2014deterministic} to optimize \eqref{eq:dpgobjective}. Specifically, given a policy network $\pi^\theta$, we wish to optimize:
\[\min_{\theta} \Expect_{\substack{\tilde{t}\sim \{0,\dots,T-1\}\\s_{t+1}\sim P(\cdot|s_t,\bar{\pi}_t(s_t))}}[ Q_{\tilde{t}}^{\pi^\theta}(s_{\tilde{t}},\pi_{\tilde{t}}^\theta(s_{\tilde{t}}))]\,,\]
using a stochastic gradient algorithm. In doing so, we rely on the fact that:
\begin{align*}
\nabla_\theta\Expect_{\substack{\tilde{t}\sim \{0,\dots,T-1\}\\s_{t+1}\sim P(\cdot|s_t,\bar{\pi}_t(s_t))}}&[ Q_{\tilde{t}}^{\pi^\theta}(s_{\tilde{t}},\pi^\theta(s_{\tilde{t}}))]\\
&=\Expect_{\substack{\tilde{t}\sim \{0,\dots,T-1\}\\s_{t+1}\sim P(\cdot|s_t,\bar{\pi}_t(s_t))}}\left[\left.\nabla_\theta Q_{\tilde{t}}^{\pi^\theta}(s_{\tilde{t}},a) \right|_{a=\pi_{\tilde{t}}^\theta(s_{\tilde{t}})}+\left.\nabla_a Q_{\tilde{t}}^{\pi^\theta}(s_{\tilde{t}},a)\nabla_\theta \pi_{\tilde{t}}^\theta(s_{\tilde{t}}) \right|_{a=\pi_{\tilde{t}}^\theta(s_{\tilde{t}})}\right]\\
&\approx \Expect_{\substack{\tilde{t}\sim \{0,\dots,T-1\}\\s_{t+1}\sim P(\cdot|s_t,\bar{\pi}_t(s_t))}}\left[\left.\nabla_a Q_{\tilde{t}}^{\pi^\theta}(s_{\tilde{t}},a)\nabla_\theta \pi_{\tilde{t}}^\theta(s_{\tilde{t}}) \right|_{a=\pi_{\tilde{t}}^\theta(s_{\tilde{t}})}\right] \,.
\end{align*}
Note that in the above equation, we have dropped the the term that depends on $\nabla_\theta Q_{\tilde{t}}^{\pi^\theta}$ as is commonly done in off-policy deterministic gradient methods and usually motivated by a result of \cite{degris2012}, who argue that this approximation preserves the set of local optima in a risk neutral setting, i.e. $\rho(\cdot):=\Expect[\cdot]$. While we do consider as an important subject of future research to extend this motivation to general recursive risk measures, our numerical experiments (see Section \ref{sec:numexp:vanilla}) will confirm empirically that the quality of this  approximation permits the identification of nearly optimal hedging policies. 

Given that we do not have access to an exact expression for $Q_{\tilde{t}}^{\pi^\theta}(s_{\tilde{t}},a)$, this operator needs to be estimated directly from the training data. Exploiting the fact that $\rho$ is a utility-based shortfall risk measure, we get that:
%\[Q_t^\pi(s_t,a_t) \in \arg\min_{q} \Expect_{s_{t+1}\sim P(\cdot|s_t,\bar{\pi}_t(s_t))}[\ell(q+r(s_t,a_t,s_{t+1})-Q_{t+1}^\pi(s_{t+1},\pi_{t+1}(s_{t+1})))\]
%\EDmodified{
\[Q_t^\pi(s_t,a_t) \in \arg\min_{q} \Expect_{s_{t+1}\sim P(\cdot|s_t,a_t)}[\ell(q+r(s_t,a_t,s_{t+1})-Q_{t+1}^\pi(s_{t+1},\pi_{t+1}(s_{t+1})))]\]%}
where $\ell(y):=(\tau\1\{y>0\}-(1-\tau)\1\{y\leq 0\})y^2$ is the score function associated to the $\tau$-expectile risk measure (see Definition \ref{def:expectile}). %\cite{RMexpectile2017}).
As explained in \cite{Shen_2014} for the case of a tabular MDP, this suggests using the following stochastic gradient step to improve each expectile estimators:
\[Q_{t}^\pi(s_t,a_t)\leftarrow Q_{t}^\pi(s_t,a_t) - \alpha\partial\ell(Q_{t}^\pi(s_{t},a_t)+r(s_t,a_t,s_{t+1})-Q_{t+1}^\pi(s_{t+1},\pi_{t+1}(s_{t+1}))\,,\]
where $\partial\ell(y):=\tau\max(0,y)-(1-\tau)\max(0,-y)$ is the subdifferential of $\ell(y)$. %\SMmodified{Using this loss function, we can implement ``recursive" expectile risk minimization. This is fairly more straightforward than if one wants to consider CVaR as the risk measure. More specifically, dynamic CVaR minimization in an AC reinforcement learning framework needs the introduction of an auxiliary variable that was introduced in equation \ref{eq:cvar} as $c$. This requires considering either an independent network for $c$, or branch out this variable from one of the critic or actor networks. The introduction of this new variable will bring difficulties in fine tuning the hyper-parameters of the model.}

In the non-tabular setting, we replace $Q_{t}^\pi(s_t,a_t)$ with two estimators: i.e. the \quoteIt{main} network $Q_{t}^\pi(s_t,a_t|\theta^Q)$ for the immediate conditional risk and the \quoteIt{target} network $Q_{t}^\pi(s_t,a_t|\theta^{Q'})$ for the next period's conditional risk. The procedure consists in iterating between a step that attempts to make the main network $Q_{t}^\pi(s_t,a_t|\theta^Q)$ a good estimator of $\rho(-r(s_t,a_t,s_{t+1})+Q_{t+1}^\pi(s_{t+1},a_{t+1}|\theta^{Q'}))$ and a step that replaces the target network $Q_{t}^\pi(s_t,a_t|\theta^{Q'})$ with a network more similar to the main one $Q_{t}^\pi(s_t,a_t|\theta^Q)$. The former is achieved, similarly as with the policy network, by searching for the optimal $\theta^Q$ according to:
\[\min_{\theta^Q} \Expect_{\substack{\tilde{t}\sim \{0,\dots,T-1\}\\s_{t+1}\sim P(\cdot|s_{t},\bar{\pi}_{t}(s_{t}))}}[\ell(Q^\pi_{\tilde{t}}(s_{\tilde{t}},\bar{\pi}_{\tilde{t}}(s_{\tilde{t}})|\theta^Q)+r(s_{\tilde{t}},\bar{\pi}_{\tilde{t}}(s_{\tilde{t}}),s_{{\tilde{t}}+1})-Q_{{\tilde{t}}+1}^\pi(s_{{\tilde{t}}+1},\pi_{{\tilde{t}}+1}(s_{{\tilde{t}}+1})|\theta^{Q'}))]\,,\]
which suggests a stochastic gradient  update of the form:
\[\theta^Q\leftarrow \theta^Q - \alpha\partial\ell(Q_{\tilde{t}}^\pi(s_{\tilde{t}},\bar{\pi}_{\tilde{t}}(s_{\tilde{t}})|\theta^Q)+r(s_{\tilde{t}},\bar{\pi}_{\tilde{t}}(s_{\tilde{t}}),s_{\tilde{t}+1})-Q_{\tilde{t}+1}^\pi(s_{\tilde{t}+1},\pi_{\tilde{t}+1}(s_{\tilde{t}+1})|\theta^{Q'}))\nabla_{\theta^Q}Q_{\tilde{t}}^\pi(s_{\tilde{t}},\bar{\pi}_{\tilde{t}}(s_{\tilde{t}})|\theta^Q)\,.\]
These two types of updates are integrated in our proposed expectile-based actor-critic deep RL (a.k.a. ACRL) algorithm described in Algorithm \ref{alg:alg_actor_critic}. \EDmodified{One may note that in each episode, the reference policy $\bar{\pi}_t$ is updated to be a perturbed version of the main policy network in order to focus the accuracy of the main critic network $Q(s,a|\theta^Q)$ value and derivatives on actions that are more likely to be produced by the main policy network. We also choose to update the target networks using convex combinations operations as is done in \cite{lillicrap2015continuous} in order to improve stability of learning.}

%\EDmodified{
\begin{algorithm}[H]
\SetAlgoLined
% \KwResult{The optimal policy $\pi_\boldsymbol{\theta}(\mathbf{S}_t | \boldsymbol{\theta})$}
Randomly initialize the main actor and critic networks' parameters  $\theta^\pi$ and $\theta^Q$\;
Initialize the target actor, $\theta^{\pi^\prime}\leftarrow \theta^{\pi}$, and critic, $\theta^{Q^\prime} \leftarrow \theta^Q$, networks\;
% {\color{blue} Set $Q_{max} = M, X_t = M$ where $M$ is a big number\;}
\For{$j=1:\#Episodes$}
{
Randomly select $t \in \{0,1,...,T-1\}$\;
Sample a minibatch of N triplets $\{(s_t^i,a_t^i,s_{t+1}^i)\}_{i=1}^N$ from $P(\cdot|s_t,\bar{\pi}_t(s_t))$, where $\bar{\pi}_t(s_t):= \pi_t(s_t|\theta^{\pi}) + \mathcal{N}(0,\sigma)$\;
Update the main critic network:
\[\theta^Q\leftarrow \theta^Q - \alpha\frac{1}{N}\sum_{i=1}^N\partial\ell(Q_{t}(s_{t}^i,a_t^i|\theta^Q)+r(s_t^i,a_t^i,s_{t+1}^i)-Q_{t+1}(s_{t+1}^i,\pi_{t+1}(s_{t+1}^i|\theta^{\pi^\prime})|\theta^{Q'}))\nabla_{\theta^Q}Q_{t}(s_{t}^i,a_t^i|\theta^Q)\,.\]
Update the main actor network:
\[\theta^\pi \leftarrow \theta^\pi - \alpha\frac{1}{N} \sum_{i=1}^N \nabla_{a} Q_t(s_t^i,a|\theta^Q)|_{ a=\pi_t(s_t^i|\theta^\pi)}\nabla_{\theta^\pi}\pi_t(s_t^i|\theta^\pi)
\]
Update the target networks:
\begin{equation}
\begin{aligned}
&\theta^{Q^\prime} \leftarrow \alpha \theta^Q + (1-\alpha)\theta^{Q^\prime}\\
&\theta^{\pi^\prime} \leftarrow \alpha \theta^{\pi} + (1-\alpha)\theta^{\pi^\prime}
\end{aligned}
\end{equation}
% {\color{blue} Compute the loss and the wealth level for each sample trajectory and update $Q_{max}, X_t$}
}
 \label{alg:alg_actor_critic}
 \caption{The actor-critic RL algorithm for the dynamic recursive expectile option hedging problem (ACRL)}
\end{algorithm}

\begin{remark}
We note that in our problem, $P(s_{t+1}|s_t,a_t)=P(s_{t+1}|s_t,a_t') = \mathbb{P}(\mathbf{S}_{t+1},\mathbf{Y}_{t+1},\sProcess_{t+1}|\mathbf{S}_{t},\mathbf{Y}_{t},\sProcess_{t})$, meaning that the action is not affecting the distribution of state in the next period. This is a direct consequence of using a translation invariant risk measure, which eliminates the need to keep track of the accumulated wealth in the set of state variables as explained in \cite{marzban2020equal} and allows the reward function to  provide an immediate signal regarding the quality of implemented actions. In the context of our deep reinforcement learning approach, we observed that convergence speed is improved in training due to this property. Furthermore, the fact that this property makes the dynamics exogenous lifts the need for keeping a replay buffer, which is also known to affect negatively convergence speed. 
%In the context of dynamic programming, reducing the state space dimension by one variable can have an important impact on numerical efficiency. However, when it comes to reinforcement learning, where the state space dimensionality is not as important, this elimination of wealth is still critical. In particular, in our option hedging problem, this would translate into transforming a zero immediate reward problem that depends only on the late reward in the very last period into a problem with non-zero immediate rewards. Our numerical results show this could highly influence the convergence speed of the model along with higher quality option hedging policies in terms of dynamic and static hedging loss. Later we demonstrate this by some numerical experiments. Furthermore, a less noticeable impact of this transformation is lifting the need for keeping a replay buffer that can cause the convergence to be slow. In fact, without having the accumulated wealth in the state space, different actions taken in different periods will not cause any change in the state variables, and any set of training samples will remain unchanged during training. Having a buffer especially in the case of basket options could slow down the computational speed of the model.}
\end{remark}

\begin{remark}
It is worth noting that while there has been a large number of DRL approaches recently proposed to address risk averse MDP using coherent risk measures, to the best of our knowledge all of those that are model-free, except for two exceptions, consider a law invariant risk measure (i.e. a static risk measure) applied on the discounted sum of total rewards (see \cite{Castro2019PracticalRM,singh20a,urpi2021risk}). Such methods therefore suffer from the issues identified in Section \ref{sec:simple_example}. The two exceptions consist of \cite{Tamar2015:PGCRM} and \cite{huang2021convergence} who propose ACRL algorithms to deal with general dynamic law-invariant coherent risk measures. While being applicable to a wider range of dynamic risk measures, the two algorithms either assume that it is possible to generate samples from a perturbed version of the underlying dynamics, or rely on training three additional neural networks (namely a state distribution reweighting network, a transition perturbation network, and a Lagrangean penalisation network) concurrently with the actor and critic networks. Furthermore, only \cite{huang2021convergence} was to this date implemented yet only tested on toy tabular problem involving 12 states and 4 actions where it produced questionable performances\footnote{At the time of writing this paper, the risk averse implementation of this algorithm was unable to recommend an optimal risk neutral policy in a deterministic setting, while the risk neutral implementation produced policies that were outperformed by risk averse ones in a stochastic setting.}. While our approach can only be used with the dynamic expectile risk measure, it offers a much simpler implementation that naturally extends DDPG to the risk averse setting. Section \ref{sec:Experiments} will present a real application of this approach on an option hedging problem involving a portfolio of 6 different assets.
\end{remark}

\section{Experimental results}\label{sec:Experiments}
In this section we provide two different sets of experiments that are run over one vanilla and one basket option. %The grid-based DP is used in the case of vanilla options to compare and confirm 
We will compare both algorithmic efficiency and quality, in terms of pricing and hedging strategies, of the dynamic risk model (\DRM{}), which employs a dynamic expectile risk measure and is solved using our new ACRL algorithm, and the static risk model (\SRM{}), which employs a static expectile measure and is solved using an AORL algorithm similar to \cite{carbonneau2021deep}. 
%Next we extend the empirical experiments to the case of basket options to demonstrate the scalability of the model. 
All experiments are done using simulated price processes of five risky assets: \APPL, \AMZN{}, \FB{}, \JPM{}, \GOOGL{}. %according to their historical prices between January 2019 and January 2021. 
The price paths are simulated using correlated Brownian motions considering the empirical mean, variance, and the correlation matrix of five reference stocks (APPL, AMZN, FB, KPM, and GOOGL) over the period that spans from January 2019 to January 2021.
%where for the correlation part the Cholesky decomposition technique is used. 
The vanilla option will be over  \APPL{} while the basket option will contain all five stocks. In both cases, the maturity of the option will be one year and the hedging portfolios will be rebalanced on a monthly basis. Table \ref{tbl:data} provides the descriptive statistics of our underlying stochastic process.

\begin{table}[h]
\begin{center}
\caption{Stock data including the mean, standard deviation, and the correlation matrix}
\begin{tabular}{l|ccccc}
&	\APPL	&	\AMZN{}	&	\FB{}	&	\JPM{}	&	\GOOGL{}	\\
\hline
$S_0$	&	78.81	&	1877.94	&	221.77	&	137.25	&	1450.16	\\
$\mu$	&	-0.0015	&	-0.0017	&	-0.0001	&	0.0006	&	-0.0004	\\
$\sigma$	&	0.0298	&	0.0243	&	0.0295	&	0.0345	&	0.0246	\\
\hline
\APPL	&	1.0000	&	0.7133	&	0.7744	&	0.5383	&	0.7680	\\
\AMZN{}	&	0.7133	&	1.0000	&	0.6903	&	0.2685	&	0.6837	\\
\FB{}	&	0.7744	&	0.6903	&	1.0000	&	0.4807	&	0.8054	\\
\JPM{}	&	0.5383	&	0.2685	&	0.4807	&	1.0000	&	0.6060	\\
\GOOGL{}	&	0.7680	&	0.6837	&	0.8054	&	0.6060	&	1.0000	\\
\end{tabular}
\label{tbl:data}
\end{center}
\end{table}

\SMmodified{In what follows, we first explain the network architecture of our ACRL model, which is composed of an actor and a critic network. Then, the training procedure of the network under the conditional risk measurement  using unconditional assessment of risk is elaborated. We also numerically demonstrate the benefit of exploiting translation invariance in an option hedging problem using RL, which is for a different purpose than what is previously shown by \cite{marzban2020equal} in a DP setting. Finally, the main numerical results of the paper is presented for pricing and hedging a vanilla and a basket option, where the advantages of having a time consistent risk measurement compared to time inconsistent approach is illustrated. In particular, we first focus on the vanilla option to show the precision of our approach by bench-marking its results against a discretized DP model and then extend the results to the case of basket options.}

\subsection{Actor and critic network architecture}

% Our implementation of the ACRL algorithm inovles two networks, one for the actor and one for the critic, as shown in Figure \ref{fig:network}. The actor network takes as input a vector of dimension $m + 1$, where $m$ is the number of assets. Since the numerical experiments are performed assuming the underlying assets of the options follow a Brownian motion process, the model only needs to consider the most recent price for each asset to make investment decisions and the time to maturity. Consequently, the input vector includes the logarithm of the price divided by the initial price, and the time remaining until maturity

% for each asset we consider 2 features, the logarithm of the price divided by the strike price, and the time remaining until maturity. The first one is different for each asset, while the second one is the same value repeated for each asset. The 2 features are shown as the third dimension of the input tensor (i.e., the number of channels). 

\begin{figure}
	\centering
	\includegraphics[scale=.45]{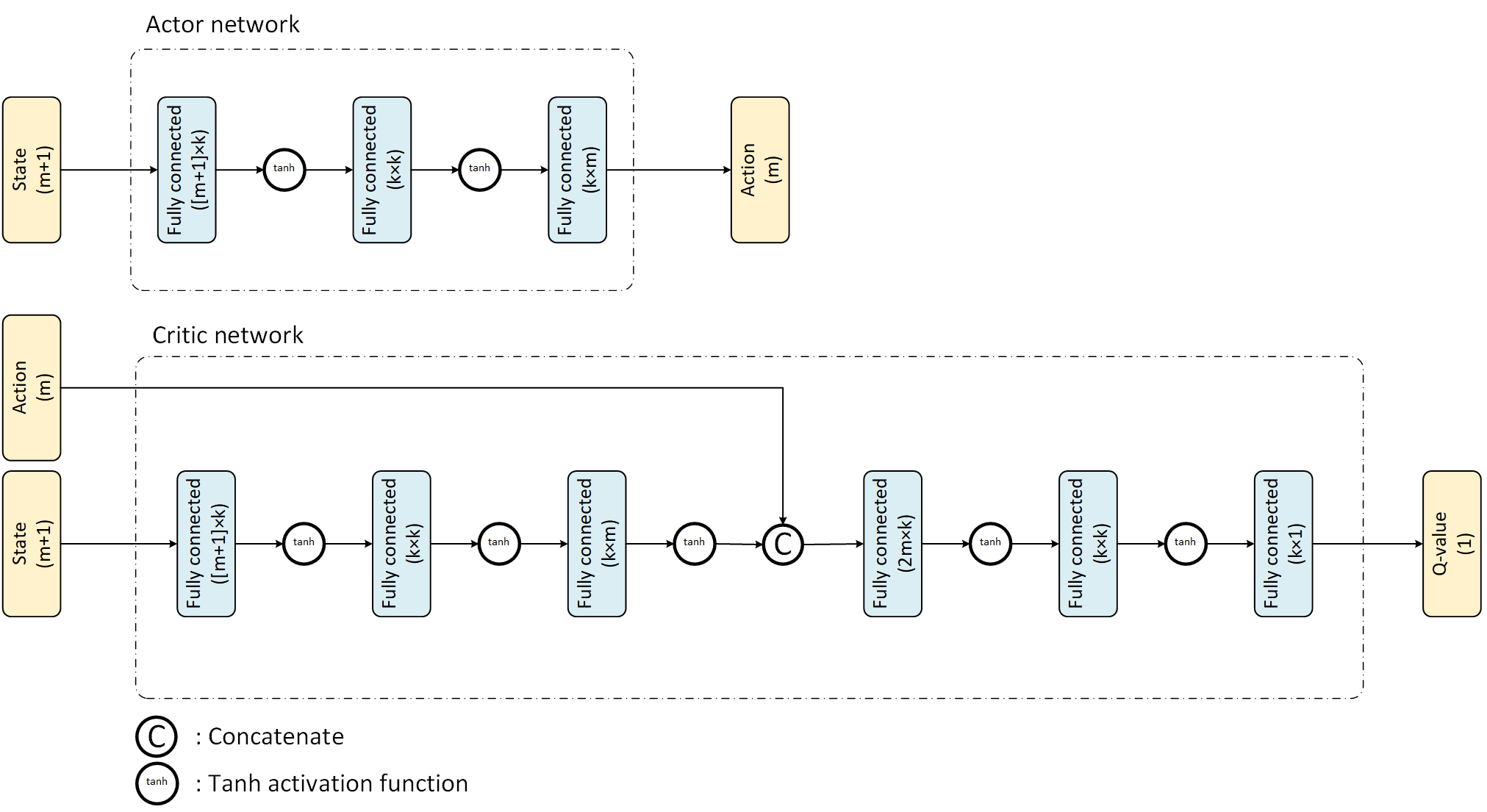}
	\caption{The architecture of the actor and critic networks in ACRL algorithm.\EDcomments{Sorry Saeed, but now I wonder if the labeling of the blue boxes would be clearer as indicating $X \rightarrow Y$ instead of $X \times Y$...}}
	\label{fig:network}
\end{figure}

Our implementation of the ACRL algorithm involves two networks, one for the actor and one for the critic, both of which are presented in Figure \ref{fig:network}. Since the numerical experiments assume that the underlying assets of the options follow a Brownian motion process, the model only needs to consider the most recent price for each asset to make investment decisions and the time to maturity. Consequently, the input state to each of the actor and critic networks includes the logarithm of each asset's cumulative return, and the time remaining until maturity, which together correspond to an input vector of dimension $m+1$.

% The actor network is composed of three fully connected layers where the kernel is of size $1 \times 1$, which ensures that the dimension of the input and output tensors stays untouched in terms of the number of assets throughout the network. However, the number of channels change to 32 after the first layer, and remains unchanged until the last layer, which reduces the number of channels to reduce to 1 to produce the investments. The activation functions in our networks are considered to be \textit{tanh}. In the last layer, this implies that the actions will lie in $a \in [-1,1]^m$.

The actor network is composed of three fully connected layers where the number of neurons are considered to be $k=32$ in the first two layers and then maps back to the number of assets in the last layer so that the model generates the investment policy accordingly for each asset. The activation functions in our networks are considered to be \textit{tanh} functions. In the last layer, this implies that the actions will lie in $[-1,\,1]^m$. 

The critic network is operating on the same state information, while the $m$ dimensional action information vector is only concatenated to the output of the third  layer. The first three layers of the critic network follow the same structure as the actor network in terms of the number of neurons, then after concatenating the action into the network, the two fully connected layers following the concatenation maps the number of neurons again to $k=32$. Finally, the last layer is a fully connected layer with one neuron to make sure that the output is a scalar representing the approximated Q value function.

\subsection{ACRL training procedure for \DRM{} and the role of translation invariance}

\begin{figure}
	\centering
	\begin{subfigure}{\mynewwidth\textwidth}
		\includegraphics[scale=\mynewscale]{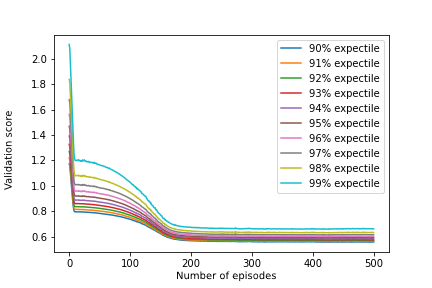}
		\caption{ACRL for {\DRM}'s writer}
	\end{subfigure}
	\begin{subfigure}{\mynewwidth\textwidth}
		\includegraphics[scale=\mynewscale]{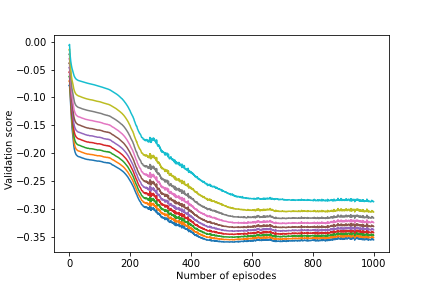}
		\caption{ACRL for {\DRM}'s buyer}
	\end{subfigure}
	
	\begin{subfigure}{\mynewwidth\textwidth}
		\includegraphics[scale=\mynewscale]{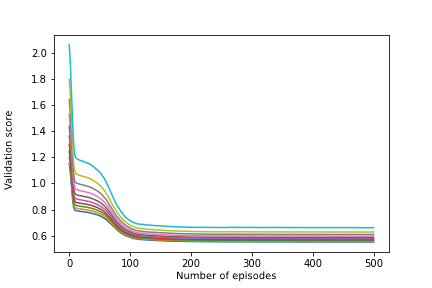}
		\caption{AORL for {\SRM}'s writer}
	\end{subfigure}
	\begin{subfigure}{\mynewwidth\textwidth}
		\includegraphics[scale=\mynewscale]{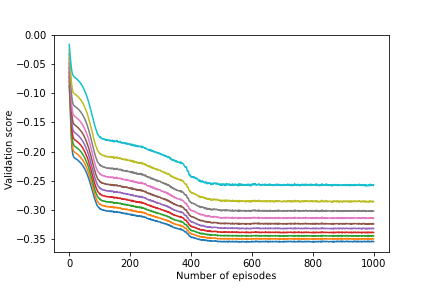}
		\caption{AORL for {\SRM}'s buyer}
	\end{subfigure}
	
	\caption{Learning curves of the \DRM{} and \SRM{} for an at-the-money vanilla call option on \APPL{} when a 90\% expectile measure is used. The graphs show the validation scores for a range of static expectile measures with risk level ranging from $90\%$ to $99\%$. \EDcomments{
y-axis label: Validation scores
Insert a legend for top left graph}\EDcomments{Validation score not validation score, i.e. capital. Still no legend ???}}
	\label{fig:vanilla_convergence}
\end{figure}

We now explain the training procedure employed for the actor and critic networks in the \DRM{}.
Recall that in an \SRM{} setting, overfitting of any DRL algorithm can be controlled by measuring the performance of the trained policy on a validation data set using an empirical estimate of the risk-averse objective as validation score. Unfortunately, this is no longer possible in the case of {\DRM}s  since the risk measure relies on conditional risk measurements of the trajectories produced by our policy. In theory, estimates of such conditional measurements could be obtained by training a new critic network using the validation set (while maintaining the policy fixed to the trained one). In practice, this is highly computationally demanding to perform in the training stage and raises a new issue of how to control overfitting of the validation score estimate. Our solution for this problem is to rely on using a static risk measure as validation score. Given that it is unclear how to best replace a dynamic expectile risk measure with a static one, we choose to compute a set of validation scores that report the performance for a set of static expectiles at risk levels that are larger or equal to the risk level of the \DRM{}. Relying on higher risk levels is motivated by the fact that dynamic expectile measures capture a more risk averse attitude than their static counterpart at the same risk level $\tau$. Figure \ref{fig:vanilla_convergence}(a) and (b) show examples of learning curves for the validation performance of a \DRM{}  when trained to hedge the writer and buyer positions of a vanilla option at a risk level of $\tau=90\%$. In this experiment, it appears that convergence roughly happens at all levels of $\tau\geq 90\%$. This approach is applied in all of our experiments for choosing the optimal number of episodes. We also note that both our training and validation sets included 1000 trajectories from the underlying geometric Brownian motion process. This implies that the training procedure used in these experiments can naturally extend to settings where only historical data is available.

%\subsection{Translation invariance and convergence}

\begin{figure}
	\centering
	\begin{subfigure}{\mynewwidth\textwidth}
		\includegraphics[scale=\mynewscale]{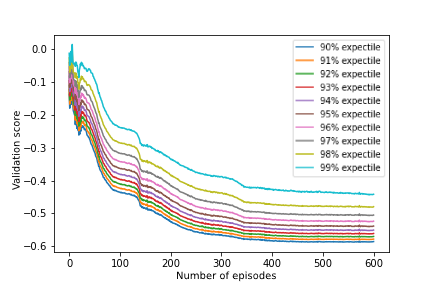}
		\caption{With immediate rewards}
	\end{subfigure}
	\begin{subfigure}{\mynewwidth\textwidth}
		\includegraphics[scale=\mynewscale]{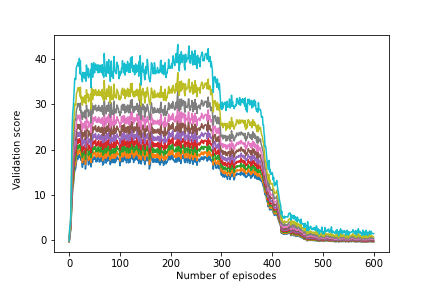}
		\caption{With delayed rewards}
	\end{subfigure}
	\caption{Learning curves of the ACRL algorithm for the buyer's \DRM{} when using (a) the immediate rewards versus (b) delayed rewards in the hedging of a vanilla call at-the-money option.\EDcomments{x-label: Number of episodes, y-label:Validation score, Insert a legend for left graph}}
	\label{fig:convergence_with_without_immediate}
\end{figure}

We close this section with a short discussion about the role of the translation invariance property of dynamic risk measures. In particular, the work of \cite{marzban2020equal} explains how without this property, the dynamic programming equations need to keep track of the wealth accumulated since $t=0$ using an additional state variable that gets only employed at $t=T$. More importantly, without translation invariance, the MDP representation ends up only having a reward at $t=T$ thus preventing the ACRL algorithm from receiving quick feedback about the quality of the actions that it is proposing. To illustrate the effect of this property, 
%
%Compared to AORL, it is well-known that the training of ACRL suffers from having two different networks that should be trained at the same time. Hyper-Parameter tuning of the ACRL algorithm is much more difficult than the AORL algorithm and it is more time consuming to train. In addition to this, the nature of option pricing problem could add up to the difficulty we are facing during training in the ACRL algorithm because of the significant delay between each action and the costs that are only observable at the maturity of the option. As discussed earlier, using translation invariant risk measures, we can transform the option hedging model to include immediate rewards into the problem of hedging an option. This section is to provide a numerical evidence on the importance of this transformation that will help easing the convergence of the ACRL algorithm. 
%
we compared the convergence of the training process for the ACRL algorithm under both form of DP representation of the buyer's \DRM{}. Namely, Figure \ref{fig:convergence_with_without_immediate} presents the learning curves of ACRL with immediate rewards as described in Section \ref{sec:ACRL}, while (b) presents the learning curves for an implementation in which all the rewards are delayed (using an additional state variable) until $t=T$. These figures clearly show that the MDP with immediate rewards is much easier to train than the delayed rewards MDP. In particular, not only does this model converge in less number of episodes, it also ends up converging to a better solution: the immediate rewards MDP converges to a risk of -0.59 for the buyer (0.91 for the writer), while with delayed rewards it converges to -0.41 (1.01).

\subsection{Vanilla call option pricing and hedging}\label{sec:numexp:vanilla}

In our first set of experiments, we consider pricing and hedging an at-the-money vanilla call option on \APPL{}. We should first note that solving a hedging problem, e.g. DRM, for a vanilla option is not particularly difficult since the number of state variables in this case is small. It is possible to obtain (approximately) optimal solutions by dynamic programming (\cite{marzban2020equal}). Our purpose of considering the case of vanilla option is twofold. First, it provides a useful basis for checking the accuracy of solutions obtained from our deep reinforcement learning (DRL) methods against the "true" optimal solutions, namely by comparing against the DP solutions. Such an accuracy check would be useful for justifying our use of DRL later in this paper as a general means to evaluate hedging performance and calculate the equal risk price (which becomes necessary for problems that cannot be solved by DP such as the case of basket options discussed in the next section). Second, the setting of a vanilla option also allows us to provide a more accurate comparison between DRM and SRM and demonstrate the advantage of the former, i.e. the benefit of time-consistent hedging policies, particularly when options with different time to maturity need to be considered.

To proceed, we first detail how the experiments are conducted. First, the initial price of the underlying stock \APPL{} is always set to be 78.81, and the hedging portfolio is rebalanced on a monthly basis. Options with different time to maturity are considered, ranging from one month to one year. We generate from a Brownian motion three sets of price trajectories with one year time window, one for training, one for validation, and one for testing, and each consists of 1000 trajectories. In the training phase, we solve both DRM and SRM for the writer and buyer's hedging problems using the longest maturity time, i.e. one year, as the hedging horizon. In solving the DRM, a policy and a critic network are trained using ACRL, whereas in solving the SRM, only a policy network is trained using AORL. See also Section 4.2 regarding how the validation is done to guide the training. Figure \ref{fig:vanilla_convergence} presents the learning curves for the training of the hedging policies of the \DRM{} and the \SRM{} with a risk level of $\tau=90\%$. \SRM{} appears to have a faster rate of convergence than \DRM{}, which might not be surprising given that the architecture of \SRM{} is simpler \footnote{The policy network at \SRM{} model is exactly the actor network of \DRM{}, while the quality of actions are directly evaluated in the absence of a trained critic network.}. It is however worth noting that the issue of time inconsistency for \SRM{} implies that it can potentially produce poor quality policies and prices when the maturity of the option is modified unless it is completely retrained for each type of maturity. This is not the case for \DRM{} and will be further discussed below.

\EDcomments{The next two paragraphs could perhaps be moved to a separate section since most of the content applies to both vanilla and basket options. This would require a bit of editing though. Let's accept it as is for now.}
With the trained DRM and SRM policy networks for a fixed 1 year maturity and risk aversion level $\tau\in\{75\%,90\%,95\%\}$, we can evaluate the writer and the buyer's (out-of-sample) risk exposure over a pre-specified time horizon so as to calculate the corresponding ERP. We consider the following three metrics for measuring the realized risk under different hedging policy and explain the methods used for calculating the metrics:
\begin{itemize}
    \item \textit{Out-of-sample static expectile risk}: Given a trained %\DRM{} and \SRM{} 
    policy network, use the testing data to calculate the static expectile risk obtained when hedging the option using this policy.
    
    \item \textit{RL based out-of-sample dynamic expectile risk estimation}: Use the testing data to train only the critic network in ACRL for evaluating the out-of-sample dynamic risk. In particular, by fixing the policy network in ACRL to a trained policy network, the critic network trained based on testing data provides an estimate of the out-of-sample dynamic expectile risk. %Thus, two critic networks are trained, one corresponding to the \DRM{} policy and the other corresponding to the \SRM{} policy. 
    To speed up the training of the critic network, one may initialize the critic network using the network trained previously with the training data.
    
    \item \textit{DP based out-of-sample dynamic expectile risk estimation}: % based on dynamic programming}: 
    Given a trained policy network, evaluate the \quoteIt{true} dynamic expectile risk by solving the dynamic programming equations, under the fixed policy, using a high precision discretization of the states, actions, and transitions. Note that this metric is available neither for the case of basket option nor in a data-driven environment where the stochastic process is unknown.
    %space and  on computing the value function over a grid from DP that employs the Brownian motion model.
    
\end{itemize}
We note that our RL based estimate of out-of-sample dynamic risk is a novel concept, which refers to the calculation of dynamic risk based on testing data. This is possible, as explained above, by training only the critic network using ACRL on the test data. This metric is especially relevant given that classical methods for calculating dynamic risk, such as our DP based estimate, assume full knowledge of the stochastic model that captures the dynamics of an underlying system, i.e. stock price, and require the resolution of dynamic programming equations, which is known to suffer from the curse of dimensionality. Consequently, such methods can no longer be used when the DP equations require a large state space, as can be the case with basket options, or when the description of the underlying stochastic process is unknown. 

In our experiments, we apply the second and third metric to the trained DRM policies and the first metric to both the trained DRM and SRM policies. % for different time horizons ranging from one month to one year (which corresponds to options with different time to maturity).
In the former case, we are interested in demonstrating that the RL based out-of-sample expectile risk estimate is an accurate metric. Namely, we will compare the RL based estimate against the \quoteIt{true} DP based estimate. In the latter case, we will shed light on how the DRM policy performs when evaluated according to other metrics that are also of interest to practitioners. In particular, the static expectile risk measure, despite its issue of time inconsistency, can still have its intuitive appeal as a metric, and one may be interested in knowing how a DRM policy performs against this metric as compared to an SRM policy. 
%Calculating different metrics for each policy not only allows one to examine how each policy would perform with respect to the risk measure that it is optimized based on, but also sheds light on how the policy would perform when evaluated according to other metrics that are also plausible to practitioners. For instance, static risk measure, despite its issue of time inconsistency, can still have its intuitive appeal as a metric, and one may be interested in knowing how a DRM policy would perform against this metric. Evaluating dynamic risk typically requires a stochastic model that captures the dynamics of an underlying system, i.e. stock price, and a dynamic programming approach, as indicated above, can be used to compute the risk. While a DP approach allows for computing the "true" risk, it relies on the availability of a stochastic model and suffers from the curse of dimensionality. The notion of out-of-sample dynamic risk proposed in our experiments is a particularly novel concept, which refers to the calculation of dynamic risk based on testing data. This is possible, as explained above, by training only the critic network using ACRL on the test data. This metric (or approach) is particularly essential when we address the case of basket option in the next section where a DP approach is not applicable. To show that out-of-sample dynamic risk can be a useful alternative metric, we will benchmark the out-of-sample dynamic risk against the true risk calculated based on DP to demonstrate the effectiveness of the former to capture dynamic risk.

\begin{figure}
	\centering
	\begin{subfigure}{\mynewwidth\textwidth}
		\includegraphics[scale=\mynewscale]{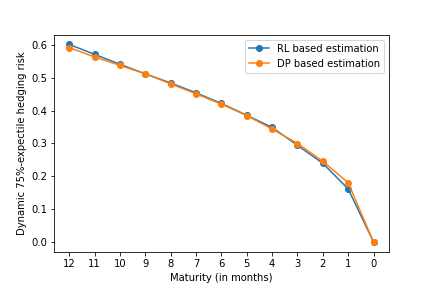}
		\caption{Writer, $\tau = 75\%$}
	\end{subfigure}
	\begin{subfigure}{\mynewwidth\textwidth}
		\includegraphics[scale=\mynewscale]{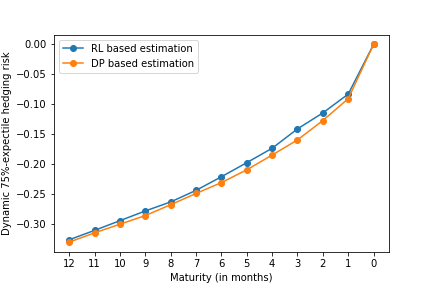}
		\caption{Buyer, $\tau = 75\%$}
	\end{subfigure}
	
	\begin{subfigure}{\mynewwidth\textwidth}
		\includegraphics[scale=\mynewscale]{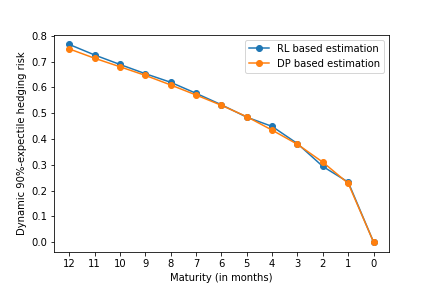}
		\caption{Writer, $\tau = 90\%$}
	\end{subfigure}
	\begin{subfigure}{\mynewwidth\textwidth}
		\includegraphics[scale=\mynewscale]{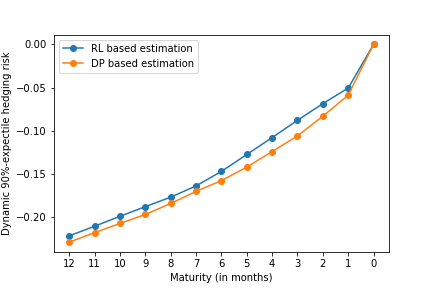}
		\caption{Buyer, $\tau = 90\%$}
	\end{subfigure}
	
	\begin{subfigure}{\mynewwidth\textwidth}
		\includegraphics[scale=\mynewscale]{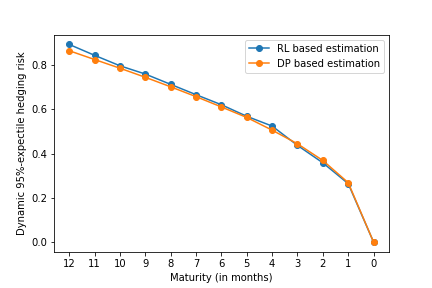}
		\caption{Writer, $\tau = 95\%$}
	\end{subfigure}
	\begin{subfigure}{\mynewwidth\textwidth}
		\includegraphics[scale=\mynewscale]{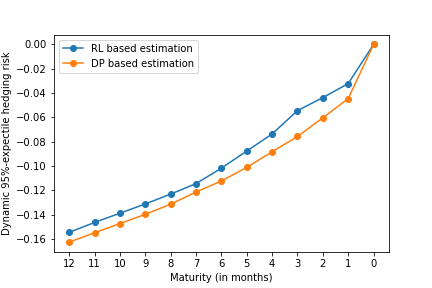}
		\caption{Buyer, $\tau = 95\%$}
	\end{subfigure}
	
	\caption{The out-of-sample dynamic risk imposed to the two sides of a vanilla at-the-money call option over \APPL{} (with maturity ranging from 12 months to 0 months) under the \DRM{} policy trained for a 12 months maturity and at different risk levels $\tau \in \{75\%,90\%,95\%\}$. %The dynamic risk is measured using either the DP value function or the Q-function.
	\EDcomments{I suggest new y-label "Dynamic XX\%-expectile hedging risk" to be clearer that graphs in different row are reporting different expectiles. Legend should be "RL based estimation", "DP based estimation"}
	\EDcomments{(e) and (f) need to be fixed regarding labels and legend.}
	\EDcomments{I also find that there is missing the ERP in these figures. I could imagine one figure per row presenting the two curves for writer and buyer with one ERP curve based on RL based measurements. In this context, I don't know if we even need the Tables.}}
	\label{fig:vanilla_loss}
\end{figure}

Figure \ref{fig:vanilla_loss} summarizes the evaluations of out-of-sample dynamic risk for DRM policies trained for 1 year maturity then applied to options of different maturities ranging from 12 months to 2 months. One can observe that the risk of the writer decreases monotonically for options of shorter maturities, whereas the risk of the buyer increases monotonically. This is consistent with the fact that there is less uncertainty for a shorter hedging horizon, which favors the writer's risk exposure more than the buyer's when considering an at-the-money option. This also provides the evidence that the DRM policies, albeit only trained based on the longest time to maturity, i.e. one year, can be well applied to hedge options with shorter time to maturity and be used to draw consistent conclusion. The observation that the DRM policies remain good policies for problems with shorter time to maturity testifies of the value of using a time consistent hedging model. %the property of time consistency of DRM policies. 
Another important observation one can make is that the RL based out-of-sample dynamic risk estimate is generally very close to the DP based estimate across all conditions. The difference between the two appears to be more noticeable for the case of high risk aversion, i.e. $\tau=95\%$ and long time to maturity, but the difference remains minor overall. This observation allows us to confirm the accuracy of our RL based out-of-sample dynamic risk estimation procedure as a replacement for the DP based estimation in settings where the latter cannot be used.

\begin{figure}
	\centering
	\begin{subfigure}{\mynewwidth\textwidth}
		\includegraphics[scale=\mynewscale]{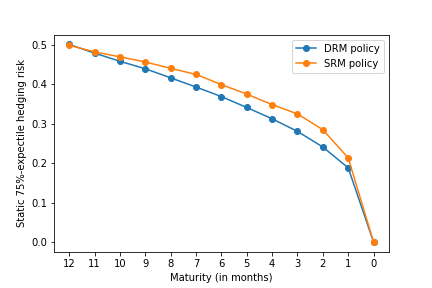}
		\caption{Writer, $\tau = 75\%$}
	\end{subfigure}
	\begin{subfigure}{\mynewwidth\textwidth}
		\includegraphics[scale=\mynewscale]{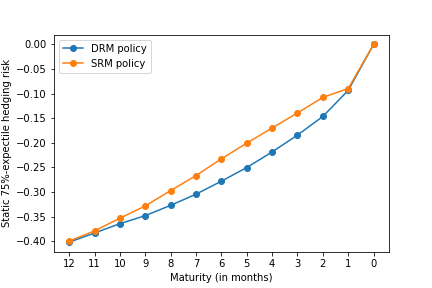}
		\caption{Buyer, $\tau = 75\%$}
	\end{subfigure}
	
	\begin{subfigure}{\mynewwidth\textwidth}
		\includegraphics[scale=\mynewscale]{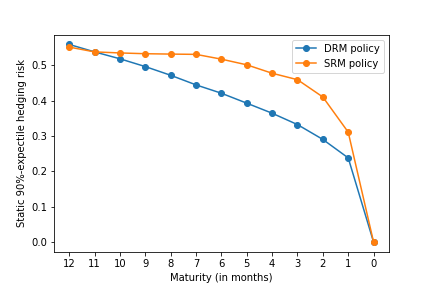}
		\caption{Writer, $\tau = 90\%$}
	\end{subfigure}
	\begin{subfigure}{\mynewwidth\textwidth}
		\includegraphics[scale=\mynewscale]{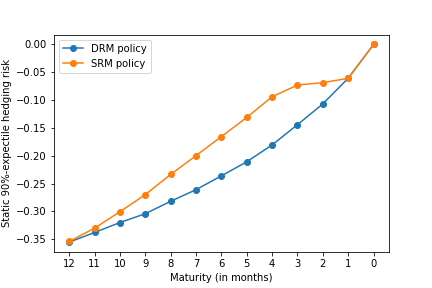}
		\caption{Buyer, $\tau = 90\%$}
	\end{subfigure}
	
	\begin{subfigure}{\mynewwidth\textwidth}
		\includegraphics[scale=\mynewscale]{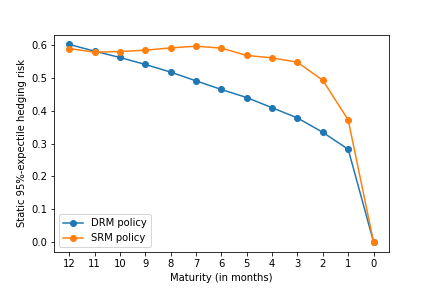}
		\caption{Writer, $\tau = 95\%$}
	\end{subfigure}
	\begin{subfigure}{\mynewwidth\textwidth}
		\includegraphics[scale=\mynewscale]{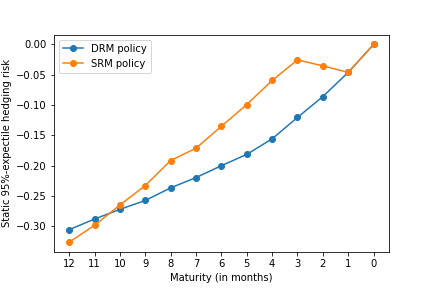}
		\caption{Buyer, $\tau = 95\%$}
	\end{subfigure}
	
	\caption{The out-of-sample static risk imposed to the two sides of a vanilla at-the-money call option over \APPL{} (with maturity ranging from 12 months to 2 months) under the \DRM{} and \SRM{} policies trained for a 12 months maturity and at different risk levels $\tau \in \{75\%,90\%,95\%\}$.\EDcomments{y-label: Static XX\%-expectile hedging risk}}
	\label{fig:vanilla_loss2}
\end{figure}

Figure \ref{fig:vanilla_loss2} reports the out-of-sample static risk for both SRM policies and DRM policies. The results are interesting and perhaps surprising. First, unlike the consistent behavior observed in the case of dynamic risk, i.e. Figure \ref{fig:vanilla_loss}, the static risk of SRM policies for the seller (resp. buyer) may increase (resp. decrease) when hedging an option with shorter maturity.
The possibility that a seller's policy may actually increase risk when applied to an option with shorter maturity is clearly problematic when the underlying asset follows a geometric brownian motion with positive drift, as it is inconsistent with the fact that there is less uncertainty (and lower expected value) regarding the payout of such options. This inconsistency occurs because the SRM policies are only trained based on the longest time to maturity, i.e. one year, and they cannot be well applied, unlike for the case of DRM policies, to problems with shorter time to maturity due to the violation of the time consistency property. It is clear from the figures that the SRM policies can be far from the optimal policies when applied to a shorter time to maturity. On the other hand, the DRM policies can actually be found not only to outperform SRM policies in terms of static risk exposure but also to generate consistent results across time, i.e. risk decreases (resp. increases) for the seller (resp. buyer) as the time to maturity decreases. This can be somewhat surprising, as the DRM policies are optimized based on dynamic risk measures rather than the static ones, but the policies can still perform well when evaluated according to static risk measures. Overall, the results presented in Figure \ref{fig:vanilla_loss2}  best showcase the strength of time consistent policies and why such policies are important to consider in settings where risk needs to be re-evaluated across different time points  or maturity dates.\footnote{Indeed, recall that the example in Section \ref{sec:simple_example} demonstrated that the fact that \SRM{} was time inconsistent implied that its policy might not remain a reasonable risk averse policy at future time points. This phenomenon is implicitly observed in Figure \ref{fig:vanilla_loss2} given that the MDP is stationary so that the risk measured for a maturity $t$ is exactly equal to the risk measured at time $T-t$ when $S_t=S_0$.} We suspect that the possibility that SRM policies may not account properly for risk aversion at some future time point or for other range of option maturities should seriously hinder their use in practice.% as the policies may simply be too aggressive to risk-averse traders.

In order to be more precise about results presented in figures \ref{fig:vanilla_loss} and \ref{fig:vanilla_loss2}, we detail in Table \ref{tbl:singleasset01} all the numerical results for the case of high risk aversion, i.e. $\tau = 90\%$ , along with the equal risk prices calculated based on RL based out-of-sample dynamic risk estimate and based on the discretized DP (referred as True ERP).\footnote{Note that in a purely data-driven setting, the ERP could either be estimated using the in-sample trained critic network, or by calculating our RL based estimate using some freshly reserved data to reduce overfitting biases.} One first confirm that the RL based estimateof ERP is a high quality approximation of the true ERP in this vanilla option pricing setting, with a maximum approximation error of 0.01 over all maturity dates. Moreover, we can see that the prices for the \SRM{} polices are generally higher than the prices for the \DRM{} polices. The observation is that while \DRM{} policies are less risky than \SRM{} policies across different time to maturity, it is the writer that benefits more from the use of \DRM{} than the buyer. This could be related to the fact that the writer's loss due to the option payout is unbounded while the option protects the buyer from losses. This in turns implies that the writer's risk exposure is larger in this transaction. Thus, the choice of a policy can be more critical to the writer than the buyer. As the risk exposure of the writer decreases more than for the buyer, this leads to lower ERP price for DRM policies.

\begin{table}[]
%\begin{center}
\small
\caption{The out-of-sample dynamic and static 90\%-expectile risk imposed to the two sides of vanilla at-the-money call options over \APPL{}, with maturities ranging from 12 to 0 months, when hedged using the \DRM{} and the \SRM{} policies trained at risk level $\tau = 90\%$ and for a 12 months maturity. Associated ERPs under the DRM are also compared to the \quoteIt{true} ERP measured using a discretized MDP. }\label{tbl:singleasset01}
\begin{tabular}{|cl|cccccccccccc|}
\hline
\multicolumn{1}{|l}{}                               &      & \multicolumn{12}{c|}{Time to maturity}                                                        \\ \hline
\multicolumn{1}{|c|}{Policy}                        & Est.$^\dagger$ & 12    & 11    & 10    & 9     & 8     & 7     & 6     & 5     & 4     & 3     & 2     & 1     \\ \hline
\multicolumn{14}{|c|}{Dynamic 90\%-expectile risk}                                                                                                         \\ \hline
\multicolumn{1}{|c|}{\multirow{2}{*}{Writer's DRM}} & RL   & 0.77  & 0.73  & 0.69  & 0.65  & 0.62  & 0.58  & 0.53  & 0.48  & 0.45  & 0.38  & 0.29  & 0.23  \\
\multicolumn{1}{|c|}{}                              & DP   & 0.75  & 0.71  & 0.68  & 0.65  & 0.61  & 0.57  & 0.53  & 0.49  & 0.43  & 0.38  & 0.31  & 0.23  \\ \hline
\multicolumn{1}{|c|}{\multirow{2}{*}{Buyer's DRM}}  & RL   & -0.22 & -0.21 & -0.20 & -0.19 & -0.18 & -0.16 & -0.15 & -0.13 & -0.11 & -0.09 & -0.07 & -0.05 \\
\multicolumn{1}{|c|}{}                              & DP   & -0.23 & -0.22 & -0.21 & -0.20 & -0.18 & -0.17 & -0.16 & -0.14 & -0.12 & -0.11 & -0.08 & -0.06 \\ \hline
\multicolumn{14}{|c|}{Static 90\%-expectile risk}                                                                                                          \\ \hline
\multicolumn{1}{|c|}{Writer's SRM}                  & ED  & 0.55  & 0.54  & 0.54  & 0.53  & 0.53  & 0.53  & 0.52  & 0.50  & 0.48  & 0.46  & 0.41  & 0.31  \\
\multicolumn{1}{|c|}{Writer's DRM}                  & ED  & 0.56  & 0.54  & 0.52  & 0.50  & 0.47  & 0.44  & 0.42  & 0.39  & 0.36  & 0.33  & 0.29  & 0.24  \\ \hline
\multicolumn{1}{|c|}{Buyer's SRM}                   & ED   & -0.35 & -0.33 & -0.30 & -0.27 & -0.23 & -0.20 & -0.17 & -0.13 & -0.09 & -0.07 & -0.07 & -0.06 \\
\multicolumn{1}{|c|}{Buyer's SRM}                   & ED   & -0.36 & -0.34 & -0.32 & -0.30 & -0.28 & -0.26 & -0.24 & -0.21 & -0.18 & -0.14 & -0.11 & -0.06 \\ \hline
\multicolumn{14}{|c|}{Equal risk prices with DRM}                                                                                                          \\ \hline
\multicolumn{2}{|c|}{True ERP}                             & 0.49  & 0.47  & 0.45  & 0.42  & 0.40  & 0.37  & 0.34  & 0.31  & 0.28  & 0.24  & 0.19  & 0.14  \\ \hline
\multicolumn{1}{|c|}{DRM}                           & RL   & 0.50  & 0.47  & 0.45  & 0.42  & 0.40  & 0.37  & 0.34  & 0.31  & 0.28  & 0.24  & 0.18  & 0.14  \\
\multicolumn{1}{|c|}{SRM}                           & RL   & 0.49  & 0.46  & 0.44  & 0.43  & 0.40  & 0.38  & 0.35  & 0.33  & 0.30  & 0.27  & 0.24  & 0.22  \\ \hline
\end{tabular}
{\footnotesize $^\dagger$ Estimation (Est.) is either made based on reinforcement learning (RL), discretized dynamic programming (DP), or with the empirical distribution (ED).}
%\end{center}
\end{table}

\begin{figure}
	\centering
	\begin{subfigure}{\mynewwidth\textwidth}
		\includegraphics[scale=\mynewscale]{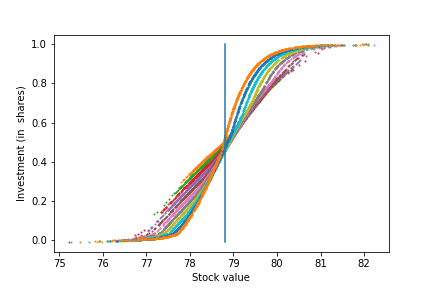}
		\caption{{\DRM}'s writer}
	\end{subfigure}
	\begin{subfigure}{\mynewwidth\textwidth}
		\includegraphics[scale=\mynewscale]{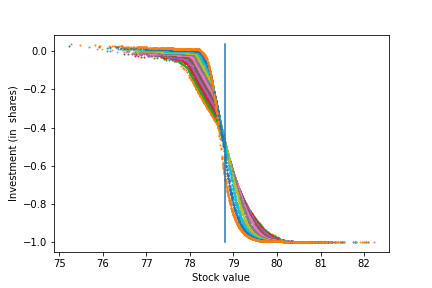}
		\caption{{\DRM}'s buyer}
	\end{subfigure}
	
	\begin{subfigure}{\mynewwidth\textwidth}
		\includegraphics[scale=\mynewscale]{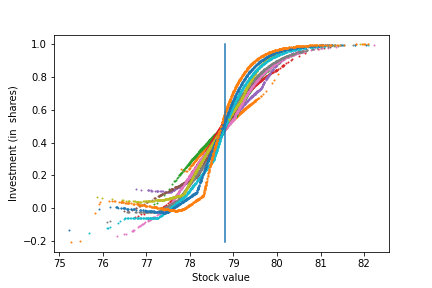}
		\caption{{\SRM}'s writer}
	\end{subfigure}
	\begin{subfigure}{\mynewwidth\textwidth}
		\includegraphics[scale=\mynewscale]{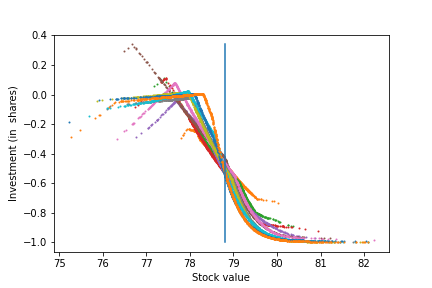}
		\caption{{\SRM}'s buyer}
	\end{subfigure}
	
	\begin{subfigure}{\mynewwidth\textwidth}
		\includegraphics[scale=\mynewscale]{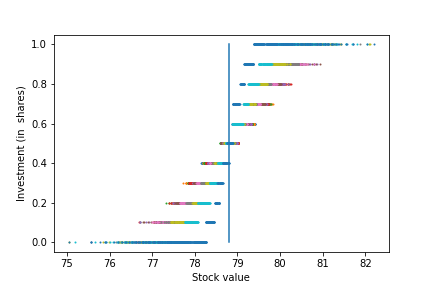}
		\caption{{\DP}'s writer}
	\end{subfigure}
	\begin{subfigure}{\mynewwidth\textwidth}
		\includegraphics[scale=\mynewscale]{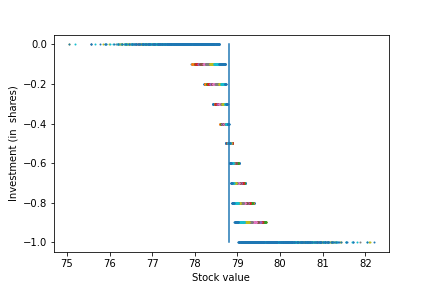}
		\caption{{\DP}'s buyer}
	\end{subfigure}
	
	\caption{Comparison of the optimal DRL policies obtained for \DRM{} and \SRM{} (with 90\% expectile measures) to the discretized DP solution (DP-DRM) for an at-the-money vanilla call option on \APPL{} with a one year maturity. Each figure presents the sampled actions in our simulated trajectories as a function of the \APPL{} stock value. The strike price is marked at 78.81. }
	\label{fig:vanilla_solution}
\end{figure}

Finally, Figure \ref{fig:vanilla_solution} presents the optimal policies of the two models (i.e., \DRM{} and \SRM{}), together with the actual optimal policy of \DRM{}, obtained using a high precision dynamic program (referred as \DP{}). Each subfigure shows the policy as a function of current price ($x$-axis) and time period (colors). The figure further confirms that the policies of both \DRM{} and \SRM{} follow a similar  pattern as \DP{}, which ensures the quality of implementation of both AORL for \SRM{} and ACRL for \DRM{}. 

%Looking more closely, a visual comparison of the optimal solutions of the \DRM{} and \SRM{} confirms the observations discussed in section \ref{sec:simple_example}. Namely, the figure shows that the \SRM{} policy is slightly less risk averse than the \DRM{} in that in can recommend to the buyer to acquire up to 0.2 share of the risky asset while both  \DRM{} and \DP{} policies mostly try to remain below 0. 
%Observing these primary results, we can now dig more precisely into the numbers generated by these models in terms of hedging loss and option prices.

\subsection{Basket options}

\begin{figure}
	\centering
	
	\begin{subfigure}{\mynewwidth\textwidth}
		\includegraphics[scale=\mynewscale]{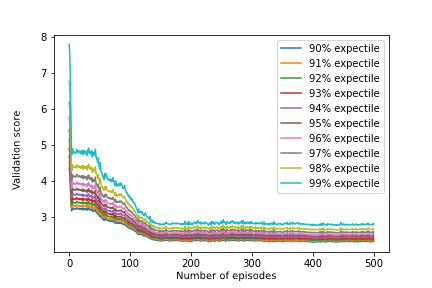}
		\caption{Writer}
	\end{subfigure}
	\begin{subfigure}{\mynewwidth\textwidth}
		\includegraphics[scale=\mynewscale]{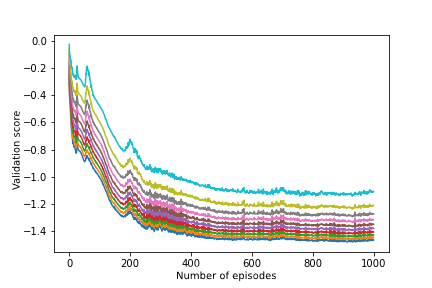}
		\caption{Buyer}
	\end{subfigure}
	
	\caption{Learning curves of the ACRL algorithm for the writer and buyer's \DRM{} for a basket at-the-money call option over \APPL{}, \AMZN{}, \FB{}, \JPM{}, and \GOOGL{} % with initial prices of 78.81, 1877.94, 221.77, 127.25, and 1450.16, and strike price of 753
 at the risk level $\tau = 90\%$. The graphs show the validation scores for a range of static expectile measures with risk level ranging from $90\%$ to $99\%$. \EDcomments{\quoteIt{Validation scores} is missing capital V. Legend missing}}
	\label{fig:basket_convergence}
\end{figure}

\begin{figure}
	\centering
	
	\begin{subfigure}{\mynewwidth\textwidth}
		\includegraphics[scale=\mynewscale]{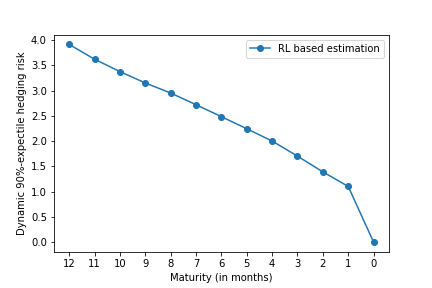}
		\caption{Writer}
	\end{subfigure}
	\begin{subfigure}{\mynewwidth\textwidth}
		\includegraphics[scale=\mynewscale]{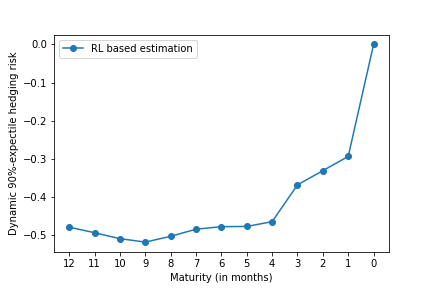}
		\caption{Buyer}
	\end{subfigure}
	
	\caption{The out-of-sample dynamic risk imposed to the two sides of a basket at-the-money call option over \APPL{}, \AMZN{}, \FB{}, \JPM{}, and \GOOGL{} at the risk level $\tau = 90\%$ (as maturity ranges from 12 to 0 months) under a \DRM{} policy trained for a 12 months maturity.}
	%The Learning curves and out-of-sample static risk imposed to the two sides of a basket at-the-money call option over \APPL{}, \AMZN{}, \FB{}, \JPM{}, and \GOOGL{} with initial prices of 78.81, 1877.94, 221.77, 127.25, and 1450.16, and strike price of 753 under the \DRM{} and the \SRM{} at the risk level $\tau = 90\%$. The \DRM{} and the \SRM{} are trained for an option with 12 month maturity and then their optimal policies are used for hedging options with shorter maturities.}
	\label{fig:basket_hedging:dynamic}
\end{figure}

\begin{figure}
	\centering
	
	\begin{subfigure}{\mynewwidth\textwidth}
		\includegraphics[scale=\mynewscale]{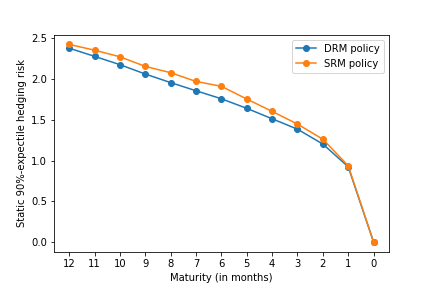}
		\caption{Static risk, writer}
	\end{subfigure}
	\begin{subfigure}{\mynewwidth\textwidth}
		\includegraphics[scale=\mynewscale]{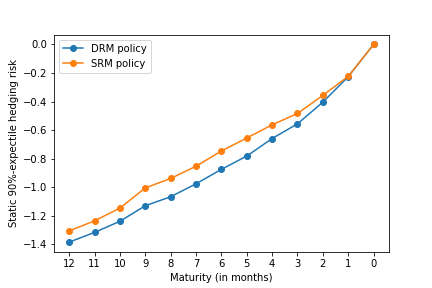}
		\caption{Static risk, buyer}
	\end{subfigure}
	
	\caption{The out-of-sample static risk imposed to the two sides of a basket at-the-money call option over \APPL{}, \AMZN{}, \FB{}, \JPM{}, and \GOOGL{} at the risk level $\tau = 90\%$ (as maturity ranges from 12 to 0 months) under the \DRM{} and \SRM{} policies trained for a 12 months maturity.}
	\label{fig:basket_hedging:static}
\end{figure}

\SMmodified{In our second set of experiments, we extend the application of ERP pricing framework to the case of basket options where traditionnal DP solution schemes are not computationally tractable. In particular, we consider an at-the-money basket option with the strike price of $753\$$ on five underlying assets: \APPL{}, \AMZN{}, \FB{}, \JPM{}, and \GOOGL{}, where the option payoff is determined by the average price of the underlyings. Similarly to the case of the vanilla option, the rebalancing of the portfolio is happening once per month, options with different maturities from one month to twelve months are considered, and three sets of price trajectories are used for training, validation, and testing the models. We train the ACRL and AORL networks for a one year basket option and then use the same policy network for hedging options with shorter time to maturity.

Our first observation in this set of experiments relates to the training time of the model for the basket option with five assets. Figure \ref{fig:basket_convergence} presents the convergence of the training of the ACRL model under $\tau = 90\%$. When comparing to the case of the vanilla option, the convergence rate appears to have a similar behavior, i.e., the number of episodes and the time spent on each episode is similar for both the case of the writer and the buyer. This is important as it indicates that the training time might not be very sensitive to the number of assets, while traditional DP approaches are known to become intractable when the option is written on multiple assets.

In this section, dynamic risk is estimated using the RL based estimator described in Section \ref{sec:numexp:vanilla} given that the DP estimator requires too much computations and that the RL based one was shown to provide a relatively high precision estimation of the ``true" dynamic risk. Following this, in Figure \ref{fig:basket_hedging:dynamic} (a) and (b) we present the dynamic risk obtained from applying the \DRM{} policy on the test data when the model is trained for a one year maturity option. Hedging risk using the same trained policy is presented for 12 different options with maturity ranging from 0 to 12 months. Similar to the vanilla option case, the dynamic risk of the writer is monotonically decreasing as we get closer to the maturity of the option, which can be attributed to the reduced probability that the average price of the  assets significantly diverges from the initial average (i.e., the strike price of the option). On the other side, i.e. for the buyer of the option, although overall the risk is increasing to zero as the maturity gets closer to zero, for longer time to maturities we observe some degradation of risk. We attribute this behavior to the estimation error of the RL based dynamic risk estimator. 

In order to have a view of risk that is not perturbed by estimation errors, we also compare the static risk under \DRM{} and \SRM{} as we did for vanilla options. Figure \ref{fig:basket_hedging:static} (a) and (b) shows the static risk under $\tau = 90\%$. One can first recognize the same monotone convergence to zero of the two sides of the options. However, contrary to the case of the vanilla option, the difference between the static risk performance of \DRM{} and \SRM{} policies are rather similar for all maturity times. It therefore appears that in these experiments with a basket option, both \SRM{} and \DRM{} produce similar polices. One possible reason could be that the range of \quoteIt{optimal} risk averse investment plans, whether using \DRM{} or \SRM{}, is more limited. Indeed, while for the vanilla option, we observed that the optimal policies generated investments in the range [0,\,1] and [-1,\,0] for the writer and the buyer respectively, for the basket option we observed wealth allocations that are more concentrated around 0.20 (i.e. the uniform portfolio known for its risk hedging properties) and -0.20 for each of the 5 assets asset respectively. 
Finally, similar to the vanilla option case, Table \ref{tbl:basketoption} presents more details on the results used to produce figures \ref{fig:basket_hedging:dynamic} and \ref{fig:basket_hedging:static}, along with the equal risk prices computed based on our RL based out-of-sample dynamic risk estimator. The higher ERP price for the \SRM{} policy is an obvious observation in this table, which again can be attributed to the better performing (in terms of dynamic risk) hedging policy produced by ACRL for the \DRM{}, compared to the policy produced by AORL for the \SRM{}.

}

\begin{table}[]
%\begin{center}
\small
\caption{The out-of-sample dynamic and static 90\%-expectile risk imposed to the two sides of basket at-the-money call options over \APPL{}, \AMZN{}, \FB{}, \JPM{}, and \GOOGL{}, with maturities ranging from 12 to 0 months, when hedged using the \DRM{} and the \SRM{} policies trained at risk level $\tau = 90\%$ and for a 12 month maturity. Associated ERPs under the DRM are also compared. \label{tbl:basketoption}}
\begin{tabular}{|cl|cccccccccccc|}
\hline
\multicolumn{1}{|l}{}              &      & \multicolumn{12}{c|}{Time to maturity}                                                        \\ \hline
\multicolumn{1}{|c|}{Policy}       & Est.$^\dagger$ & 12    & 11    & 10    & 9     & 8     & 7     & 6     & 5     & 4     & 3     & 2     & 1     \\ \hline
\multicolumn{14}{|c|}{Dynamic 90\%-expectile risk}                                                                                        \\ \hline
\multicolumn{1}{|c|}{Writer's DRM} & RL   & 3.92  & 3.62  & 3.38  & 3.15  & 2.95  & 2.72  & 2.48  & 2.25  & 2.00  & 1.70  & 1.39  & 1.10  \\ \hline
\multicolumn{1}{|c|}{Buyer's DRM}  & RL   & -0.48 & -0.49 & -0.51 & -0.52 & -0.50 & -0.49 & -0.48 & -0.48 & -0.47 & -0.37 & -0.33 & -0.29 \\ \hline
\multicolumn{14}{|c|}{Static 90\%-expectile risk}                                                                                         \\ \hline
\multicolumn{1}{|c|}{Writer's SRM} & ED   & 2.43  & 2.36  & 2.28  & 2.16  & 2.08  & 1.97  & 1.91  & 1.76  & 1.61  & 1.45  & 1.26  & 0.94  \\
\multicolumn{1}{|c|}{Writer's DRM} & ED   & 2.38  & 2.28  & 2.18  & 2.06  & 1.96  & 1.86  & 1.76  & 1.64  & 1.51  & 1.39  & 1.20  & 0.92  \\ \hline
\multicolumn{1}{|c|}{Buyer's SRM}  & ED   & -1.31 & -1.24 & -1.15 & -1.01 & -0.94 & -0.85 & -0.75 & -0.66 & -0.56 & -0.48 & -0.36 & -0.22 \\
\multicolumn{1}{|c|}{Buyer's SRM}  & ED   & -1.39 & -1.32 & -1.24 & -1.13 & -1.07 & -0.98 & -0.88 & -0.78 & -0.66 & -0.56 & -0.40 & -0.23 \\ \hline
\multicolumn{14}{|c|}{Equal risk prices with DRM}                                                                                         \\ \hline
\multicolumn{1}{|c|}{DRM}          & RL   & 2.20  & 2.06  & 1.95  & 1.84  & 1.73  & 1.61  & 1.48  & 1.37  & 1.24  & 1.04  & 0.86  & 0.70  \\
\multicolumn{1}{|c|}{SRM}          & RL   & 2.23  & 2.10  & 2.01  & 1.91  & 1.79  & 1.65  & 1.52  & 1.39  & 1.21  & 1.03  & 0.92  & 0.82  \\ \hline
\end{tabular}
%\end{center}
{\footnotesize $^\dagger$ Estimation (Est.) is either made based on reinforcement learning (RL), discretized dynamic programming (DP), or with the empirical distribution (ED).}
\end{table}

\section{Conclusion}

In this paper, we developed and implemented the first deep reinforcement learning algorithm for calculating equal risk prices under time consistent dynamic risk measures. This algorithm exploits the elicitability property of the expectile risk measure to extend in a natural way the famous off-policy deterministic actor-critic method presented in \cite{silver2014deterministic} to the risk averse setting. Our numerical experiments confirmed that it can identify risk averse hedging strategies of good quality and be used to estimate the ERP, simultaneously for a range of maturities, using a reasonable amount of computational resources in conditions where traditional DP methods are impracticable. We also demonstrated important issues regarding the implementability of hedging strategies that are based on static (time inconsistent) risk measures. Namely, both our illustrative example and two numerical experiments demonstrated how the time consistent policy produced using the DRM might in fact appear preferable to the investor (from the point of view of the time inconsistent static risk measure) as the risk is measured at later points of time, i.e. with shorter maturity. \EDcomments{Is there more to say about this issue? What are natural extension (empirical study on real date) directions for future work (extending the Degris result) ?} \SMmodified{Overall, as the first paper that is considering option pricing under ERP using time consistent dynamic risk measures, we only evaluated the performance of our model in a synthetic environment using a simple neural network architecture. One may be interested in examining the performance of this model under real market conditions. In particular, in a simulation environment having access to infinite i.i.d. samples makes training much easier to machine learning models. In a real market environment, where the available data is limited to some non-stationary samples of past historical prices, training an on-policy network will face serious issues associated to lack of exploration. This might result in even higher out performance of the ACRL model compared to the AORL, and therefore superior hedging precision under the time consistent dynamic risk measure. In addition, we only consider European style options in this paper, where as demonstrated in \cite{marzban2020equal}, the ERP model can also be investigated in the case of American options.}

\bibliographystyle{unsrtnat}
\bibliography{References}

\begin{thebibliography}{21}
\providecommand{\natexlab}[1]{#1}
\providecommand{\url}[1]{\texttt{#1}}
\expandafter\ifx\csname urlstyle\endcsname\relax
  \providecommand{\doi}[1]{doi: #1}\else
  \providecommand{\doi}{doi: \begingroup \urlstyle{rm}\Url}\fi

\bibitem[Guo and Zhu(2017)]{guo2017equal}
Ivan Guo and Song-Ping Zhu.
\newblock Equal risk pricing under convex trading constraints.
\newblock \emph{Journal of Economic Dynamics and Control}, 76:\penalty0
  136--151, 2017.

\bibitem[Marzban et~al.(2020)Marzban, Delage, and Li]{marzban2020equal}
Saeed Marzban, Erick Delage, and Jonathan~Yumeng Li.
\newblock Equal risk pricing and hedging of financial derivatives with convex
  risk measures.
\newblock \emph{arXiv preprint arXiv:2002.02876}, 2020.

\bibitem[Carbonneau and Godin(2020)]{carbonneau2020equal}
Alexandre Carbonneau and Fr{\'e}d{\'e}ric Godin.
\newblock Equal risk pricing of derivatives with deep hedging.
\newblock \emph{Quantitative Finance}, pages 1--16, 2020.

\bibitem[Carbonneau and Godin(2021)]{carbonneau2021deep}
Alexandre Carbonneau and Fr{\'e}d{\'e}ric Godin.
\newblock Deep equal risk pricing of financial derivatives with multiple
  hedging instruments.
\newblock \emph{arXiv preprint arXiv:2102.12694}, 2021.

\bibitem[Williams(1992)]{williams1992simple}
Ronald~J Williams.
\newblock Simple statistical gradient-following algorithms for connectionist
  reinforcement learning.
\newblock \emph{Machine learning}, 8\penalty0 (3):\penalty0 229--256, 1992.

\bibitem[Mnih et~al.(2015)Mnih, Kavukcuoglu, Silver, Rusu, Veness, Bellemare,
  Graves, Riedmiller, Fidjeland, Ostrovski, et~al.]{mnih2015human}
Volodymyr Mnih, Koray Kavukcuoglu, David Silver, Andrei~A Rusu, Joel Veness,
  Marc~G Bellemare, Alex Graves, Martin Riedmiller, Andreas~K Fidjeland, Georg
  Ostrovski, et~al.
\newblock Human-level control through deep reinforcement learning.
\newblock \emph{Nature}, 518\penalty0 (7540):\penalty0 529--533, 2015.

\bibitem[Silver et~al.(2014)Silver, Lever, Heess, Degris, Wierstra, and
  Riedmiller]{silver2014deterministic}
David Silver, Guy Lever, Nicolas Heess, Thomas Degris, Daan Wierstra, and
  Martin Riedmiller.
\newblock Deterministic policy gradient algorithms.
\newblock In \emph{International conference on machine learning}, pages
  387--395. PMLR, 2014.

\bibitem[Tamar et~al.(2015)Tamar, Chow, Ghavamzadeh, and
  Mannor]{Tamar2015:PGCRM}
Aviv Tamar, Yinlam Chow, Mohammad Ghavamzadeh, and Shie Mannor.
\newblock Policy gradient for coherent risk measures.
\newblock In C.~Cortes, N.~Lawrence, D.~Lee, M.~Sugiyama, and R.~Garnett,
  editors, \emph{Advances in Neural Information Processing Systems}, volume~28.
  Curran Associates, Inc., 2015.

\bibitem[Huang et~al.(2021)Huang, Leqi, Lipton, and
  Azizzadenesheli]{huang2021convergence}
Audrey Huang, Liu Leqi, Zachary~C. Lipton, and Kamyar Azizzadenesheli.
\newblock On the convergence and optimality of policy gradient for markov
  coherent risk, 2021.

\bibitem[Bertsimas et~al.(2001)Bertsimas, Kogan, and Lo]{bertsimas2001hedging}
Dimitris Bertsimas, Leonid Kogan, and Andrew~W Lo.
\newblock Hedging derivative securities and incomplete markets: an
  $\epsilon$-arbitrage approach.
\newblock \emph{Operations research}, 49\penalty0 (3):\penalty0 372--397, 2001.

\bibitem[Artzner et~al.(1999)Artzner, Delbaen, Eber, and
  Heath]{artzner1999coherent}
Philippe Artzner, Freddy Delbaen, Jean-Marc Eber, and David Heath.
\newblock Coherent measures of risk.
\newblock \emph{Mathematical finance}, 9\penalty0 (3):\penalty0 203--228, 1999.

\bibitem[Rudloff et~al.(2014)Rudloff, Street, and
  Vallad{\~a}o]{rudloff2014time}
Birgit Rudloff, Alexandre Street, and Davi~M Vallad{\~a}o.
\newblock Time consistency and risk averse dynamic decision models: Definition,
  interpretation and practical consequences.
\newblock \emph{European Journal of Operational Research}, 234\penalty0
  (3):\penalty0 743--750, 2014.

\bibitem[Bellini and Bignozzi(2015)]{bellini2015:elicitable}
Fabio Bellini and Valeria Bignozzi.
\newblock On elicitable risk measures.
\newblock \emph{Quantitative Finance}, 15\penalty0 (5):\penalty0 725--733,
  2015.

\bibitem[Chen(2018)]{chen2018exactitude}
James~Ming Chen.
\newblock On exactitude in financial regulation: Value-at-risk, expected
  shortfall, and expectiles.
\newblock \emph{Risks}, 6\penalty0 (2):\penalty0 61, 2018.

\bibitem[Bellini and Bernardino(2017)]{RMexpectile2017}
Fabio Bellini and Elena~Di Bernardino.
\newblock Risk management with expectiles.
\newblock \emph{The European Journal of Finance}, 23\penalty0 (6):\penalty0
  487--506, 2017.

\bibitem[Degris et~al.(2012)Degris, White, and Sutton]{degris2012}
Thomas Degris, Martha White, and Richard~S. Sutton.
\newblock Off-policy actor-critic.
\newblock In \emph{Proceedings of the 29th International Coference on
  International Conference on Machine Learning}, ICML'12, page 179–186,
  Madison, WI, USA, 2012. Omnipress.

\bibitem[Shen et~al.(2014)Shen, Tobia, Sommer, and Obermayer]{Shen_2014}
Yun Shen, Michael~J. Tobia, Tobias Sommer, and Klaus Obermayer.
\newblock Risk-sensitive reinforcement learning.
\newblock \emph{Neural Computation}, 26\penalty0 (7):\penalty0 1298–1328,
  2014.

\bibitem[Lillicrap et~al.(2015)Lillicrap, Hunt, Pritzel, Heess, Erez, Tassa,
  Silver, and Wierstra]{lillicrap2015continuous}
Timothy~P Lillicrap, Jonathan~J Hunt, Alexander Pritzel, Nicolas Heess, Tom
  Erez, Yuval Tassa, David Silver, and Daan Wierstra.
\newblock Continuous control with deep reinforcement learning.
\newblock \emph{arXiv preprint arXiv:1509.02971}, 2015.

\bibitem[Castro et~al.(2019)Castro, Oren, and Mannor]{Castro2019PracticalRM}
Dotan~Di Castro, J.~Oren, and Shie Mannor.
\newblock Practical risk measures in reinforcement learning.
\newblock \emph{ArXiv}, abs/1908.08379, 2019.

\bibitem[Singh et~al.(2020)Singh, Zhang, and Chen]{singh20a}
Rahul Singh, Qinsheng Zhang, and Yongxin Chen.
\newblock Improving robustness via risk averse distributional reinforcement
  learning.
\newblock In Alexandre~M. Bayen, Ali Jadbabaie, George Pappas, Pablo~A.
  Parrilo, Benjamin Recht, Claire Tomlin, and Melanie Zeilinger, editors,
  \emph{Proceedings of the 2nd Conference on Learning for Dynamics and
  Control}, volume 120 of \emph{Proceedings of Machine Learning Research},
  pages 958--968, 2020.

\bibitem[{Urpí} et~al.(2021){Urpí}, {Curi}, and {Krause}]{urpi2021risk}
Núria~Armengol {Urpí}, Sebastian {Curi}, and Andreas {Krause}.
\newblock Risk-averse offline reinforcement learning.
\newblock In \emph{ICLR 2021: The Ninth International Conference on Learning
  Representations}, 2021.

\end{thebibliography}
\end{document}